\documentclass[twocolumn,showpacs,preprintnumbers,amsmath,amssymb,epsf,superscriptaddress]{revtex4-1}

\usepackage{graphicx,epsfig}
\usepackage{dcolumn}
\usepackage{bm}

\begin{document}


\title{Emergence and seismological implications of phase transition and universality in a system with interaction between thermal pressurization and dilatancy}
\author{Takehito Suzuki}
\email{t-suzuki@phys.aoyama.ac.jp}
\affiliation{Department of Physics and Mathematics, Aoyama Gakuin University, 5-10-1 Fuchinobe, Sagamihara 252-5258, Japan}


\begin{abstract}
A dynamic earthquake source process is modeled by assuming interaction among frictional heat, fluid pressure, and inelastic porosity. In particular, fluid pressure increase due to frictional heating (thermal pressurization effect) and fluid pressure decrease due to inelastic porosity increase (dilatancy effect) play important roles in this process. Two nullclines become exactly the same in the system of governing equations, which generates non-isolated fixed points in the phase space. These lead to a type of phase transition, which produces a universality described by the power law between the initial value of one variable and the final value of the other variable. The universal critical exponent is found to be $1/2$, which is independent of the details of the porosity evolution law. We can regard the dynamic earthquake slip process as a phase transition by considering the final porosity or slip as the order parameter. Physical prediction of phase emergence is difficult because the porosity evolution law has uncertainties, and the final slip amount is difficult to predict because of the universality. Finally, nonlinear mathematical application of the result is also discussed.
\end{abstract}

\maketitle

\section{Introduction} \label{secI}
Dynamic earthquake slip processes are frictional slips inside the Earth. They show various qualitatively different behaviors, and discontinuous transitions between them are widely observed. For example, several earthquakes show pulse-like slips, whereas some generate crack-like slips as their slip profiles \cite{Amp}. The existence of slow earthquakes with slip velocity and rupture velocity (the propagation velocity of the fault tip) much smaller than those for ordinary earthquakes is widely known \cite{Oba}. There are also a variety of stress drops, which is defined as the difference between the shear stress acting on the fault plane and the residual shear stress. Almost all earthquakes have scale-independent stress drops; however, some are considered to have enormously large ones \cite{Dup}. Whether intermediate cases exist or not is a controversial problem. If we refer to each aspect as a phase, these behaviors may be understood in terms of phase transitions in a unified manner.

To explain such transitions, it is insufficient to assume cracks in classical elastic bodies. Several researchers have noticed several aspects of the interior of the Earth, such as fault rock melting \cite{Spr} and chemical effects \cite{DiT}. For example, the slow slip velocity and rupture velocity for slow earthquakes are impossible to reproduce with the classical crack model; generation of slow earthquakes is considered to be promoted by certain factors, e.g., migration of the fluid (water) \cite{Suz14, Seg}. Some studies have treated interactions among these effects. For example, frictional heating and fluid pressure are considered to interact as thermal pressurization \cite{Lac}. This interaction describes elevation of fluid pressure due to frictional heating. Such fluid pressure elevation induces a decrease in the normal stress acting on the fault plane, leading to frictional stress decrease and slip acceleration. Many studies, including a sequence of studies by the author, have considered another interaction between thermal pressurization and dilatancy (referred to as ITPD below) \cite{Suz14, Seg, Ric06, Suz07, Suz08, Suz09, Suz10}. The dilatancy is slip-induced inelastic pore creation, leading to fluid pressure decrease and slip deceleration. This interaction enables us to treat slip acceleration and deceleration in a single framework, with which we can understand both ordinary and slow earthquakes.

However, some problems remain unsolved in the treatment of ITPD. For example, it should be emphasized that the porosity evolution laws are not yet firmly understood \cite{Bart, Teu, Rud, Mar, Sle}. Laboratory experiments under conditions deep in the Earth are so hard to perform that exact reproduction of porosity evolution is considered difficult. A unified treatment independent of the details of the law is therefore required to understand the behavior of the system in the presence of ITPD. Additionally, geophysical studies have solely focused on explaining geophysical phenomena, and mathematically and physically important characteristics have not been investigated, as indicated below.

The studies of ITPD mentioned above employed nonlinear governing equation systems. In fact, the behaviors of solution orbits for nonlinear equation systems have been studied widely. The competitive Lotka-Volterra (LV) equation system, a model describing competition between two species, is an example of such a nonlinear equation system and has been a frequently treated topic recently \cite{Chen, Fer, Fel}. For such a system, isolated fixed points on the phase space have been found, and the features of solution orbits crossing the points are well understood. The points are attractors, a saddle node and a repeller: the attractors describe extinction of one species, the saddle node corresponds to coexistence of both species, and the origin is the repeller. Discontinuous change of the solution behaviors depending on the initial value is observed there, similar to the discontinuous behaviors observed in the dynamic earthquake slip process. Nonetheless, two characteristics need to be emphasized in the ITPD model. First, continuous non-isolated fixed points appear for the ITPD model as attractors, as shown in this paper. Second, the initial values of the variables construct the continuous geometry in the ITPD model; the group of the initial values becomes a line in the phase space. We can therefore conclude that the initial and final values construct the continuous geometries in the ITPD model. We can expect a universal relation between these continuous geometries, although constructing the framework to treat such a relation has not been achieved.

This paper is organized as follows. The model setup and governing equation system are clarified in Sec. \ref{secMS}. The slip velocity and inelastic porosity are the variables governing system behavior. Geometrically different attractors generated by the nullclines common to both variables are found mathematically in Sec. \ref{secATT}. Discontinuity of the solution behaviors can be regarded as phase transition behavior. The criticality in the vicinity of the phase transition point is found in Sec. \ref{secPT}. The criticality is universal, and is unaffected by the assumed details of the porosity evolution law. Physical and seismological application of the results obtained is carried out in Sec. \ref{secApp}. Dynamic earthquake slip behavior is concluded to be the phase transition phenomenon. In particular, predicting which phase emerges and the extent of the final slip amount is difficult. The paper is summarized, and nonlinear mathematical application is performed in Sec. \ref{secDisCon}.

\section{MODEL SETUP} \label{secMS}

\begin{figure}[tbp]
\centering
\includegraphics[width=8.5cm]{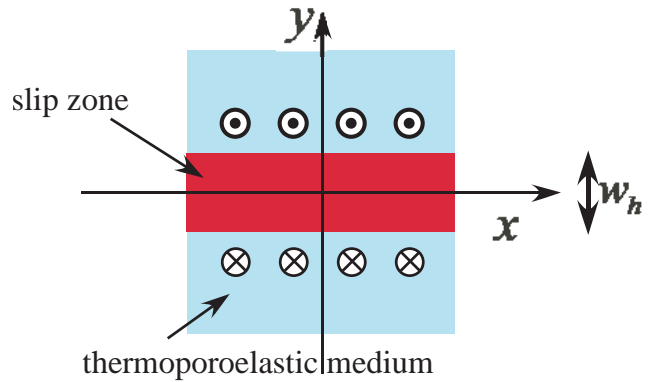}
\caption{Model setup. The thermoporoelastic medium moves as symbols show. See also Figure 1 in SY14.}
\label{FigMS}
\end{figure}

We consider a system consisting of a homogeneous and isotropic thermoporoelastic medium, i.e. the medium has pores whose volume ratio to the whole volume (porosity) is initially homogeneous. The initial porosity is referred to as elastic porosity, $\phi_e$. The pores are assumed to be filled with fluid (water). We also assume that the thermal pressurization and dilatancy effects emerge in the slip zone located at $-w_h/2 < y < w_h/2$ along the $x$-axis (Fig. \ref{FigMS}); the relative movement between opposite surfaces is assumed to be accommodated entirely within the slip zone, which is regarded macroscopically as the one-dimensional (1D) mode III slip plane. The slip zone can be considered as a boundary of the medium from a macroscopic viewpoint. Thermal pressurization describes the elevation in fluid pressure based on the frictional heating associated with the dynamic fault slip \cite{Lac}. If frictional slip occurs, frictional heating increases the temperature, inducing expansion of the solid and fluid phases. However, because it is easier to expand the fluid phase than the solid phase, the fluid pressure rises. On the other hand, dilatancy represents the inelastic porosity increase due to the fracturing of fault rocks by the fault slip, which reduces the fluid pressure \cite{Seg}. It should be emphasized that when the fluid pressure increases (decreases), the frictional stress decreases (increases) due to a reduction (increment) in the effective normal stress acting on the fault plane, inducing the slip velocity increase (decrease). The competition between thermal pressurization and dilatancy induces complex feedback in the slip behavior, which can explain many aspects of the dynamic earthquake slip process (e.g., Suzuki and Yamashita (2014) \cite{Suz14}, referred to as SY14 below). However, this system is not yet researched from the viewpoint of nonlinear mathematics, particularly with regard to the behaviors of attractors. This research therefore primarily aims at understanding the system with regard to the behaviors of attractors.

We have the following system of governing equations within the slip zone (SY14):
\begin{equation}
\frac{1}{M} \frac{\partial p_D}{\partial t} = ((b-\phi_t) \alpha_s + \phi_t \alpha_f) \frac{\partial T_D}{\partial t} -\frac{\partial \phi_d}{\partial t}, \label{eqP}
\end{equation}
\begin{equation}
((1-\phi_t) \rho_s C_s + \phi_t \rho_f C_f) \frac{\partial T_D}{\partial t} = \frac{\sigma_{\mathrm{res}} v_D}{w_h}, \label{eqT} 
\end{equation}
\begin{equation}
v_D=\frac{2 \beta_v}{\mu} (\sigma_s^0 -\sigma_{\mathrm{res}}), \label{eqEOM}
\end{equation}
\begin{equation}
\sigma_{\mathrm{res}}=-\mu_{\mathrm{slid}} (\sigma_n^0+p_D). \label{eqSres}
\end{equation}
Table I summarizes the meanings of the parameters. We neglected the advection effect and adiabatic expansion of the solid phase; the validity of this assumption has been demonstrated in our previous study \cite{Suz10}. We also neglected the diffusions of the fluid and heat, such that the Heaviside unit step function used in SY14 is not required and variables do not depend on $y$ because the slip zone becomes an isolated system with such assumptions.

\begin{table*}{}
\caption{Properties, their meanings and values. Values are based on SY14. However, the values are solely used for estimating $t^f_{\mathrm{ch}}$ and $t^h_{\mathrm{ch}}$, and are not adopted in the normalized equation system.}
\begin{tabular}{lll}
\hline\hline
Properties & Physical Meanings & Values \\ \hline
$b=1-\frac{K_v}{K_s} $ & & 0.2 \\
$C_s$ & Specific heat for the solid phase & $9.2 \times 10^2 \ \mathrm{J} \ \mathrm{kg}^{-1} \ \mathrm{K}^{-1}$ \\
$C_f$ & Specific heat for the fluid phase & $4.2 \times 10^3 \ \mathrm{J} \ \mathrm{kg}^{-1} \ \mathrm{K}^{-1}$ \\
$K_s$ & Bulk modulus of the solid phase & $3 \times 10^4$ MPa \\
$K_f$ & Bulk modulus of the fluid phase & $3.3 \times 10^3$ MPa \\
$K_v$ & Bulk modulus of the medium & $2.4 \times 10^4$ MPa \\
$M=\left( \frac{b-\phi_t}{K_s} +\frac{\phi_t}{K_f} \right)^{-1}$ & & $2.97 \times 10^4$ MPa \\
$p_D$ & Fluid pressure & -\footnotemark[1] \\
$T_D$ & Temperature & -\footnotemark[1] \\
$u_D$ & Slip & -\footnotemark[1] \\
$v_D$ & Slip velocity & -\footnotemark[1] \\
$w_h$ & Slip zone thickness & 3 mm to 3 cm \\
$\alpha_s$ & Thermal expansion coefficient of the solid phase & $1 \times 10^{-5}$ $\mathrm{K}^{-1}$ \\
$\alpha_f$ & Thermal expansion coefficient of the fluid phase & $2.1 \times 10^{-4}$ $\mathrm{K}^{-1}$ \\
$\beta_v=\sqrt{\frac{\mu}{(1-\phi_t)\rho_s +\phi_t \rho_f}}$ & Shear wave speed & $2.39 \times 10^3 \ \mathrm{m} \ \mathrm{s}^{-1}$ \\
$\eta$ & Fluid phase viscosity & $2.82 \times 10^{-4}$ Pa s \\
$\mu$ & Shear modulus of the medium & $1.44 \times 10^4$ MPa \\
$\mu_{\mathrm{slid}}$ & Sliding frictional coefficient & 0.6 \\
$\rho_s$ & Solid phase density & $2.7 \times 10^3 \ \mathrm{kg} \ \mathrm{m}^{-3}$ \\
$\rho_f$ & Fluid phase density & $1 \times 10^3 \ \mathrm{kg} \ \mathrm{m}^{-3}$ \\
$\sigma_n^0$ & Normal stress acting on the fault & $-2.5 \times 10^2$ MPa \\
$\sigma_s^0$ & Shear stress acting on the fault & $1 \times 10^2$ MPa \\
$\sigma_{\mathrm{res}}$ & Residual frictional stress & -\footnotemark[1] \\
$\phi_d$ & Inelastic porosity & -\footnotemark[1] \\
$\phi_e$ & Elastic porosity & 0.1 \\
$\phi_t$ & Total porosity ($=\phi_d+\phi_e$) & Assumed to be 0.1 \\ \hline\hline
\footnotetext[1]{Functions of time}
\end{tabular}
\end{table*}

We now consider the characteristic fluid diffusion time, $t^f_{\mathrm{ch}}$, and roughly estimate it here to show that neglecting fluid diffusion can be valid in the interior of the Earth. For the estimation, we should first mention that the slip zone thickness is known to be on the order of $\mu$m-cm \cite{Ric06, Ches, Hee, Sib, Uji07, Uji08, Kam, Row, Pla, Ric14}. For subduction thrust faults, the order of mm-cm seems to be reasonable \cite{Uji07, Uji08, Kam, Row}. On the other hand, some researchers insist that the zones have width on the order of $\mu$m \cite{Pla, Ric14}. However, significant along-strike variability may be observed in the localized zone thickness \cite{Ric14}, and the fault zones accommodating displacements are considered to have complex geometrical structures \cite{Ben}. We therefore assume an order of a few cm or less in this study. Additionally, the permeability $k$ is known to be on the order of  $10^{-14}$-$10^{-21} \ \mathrm{m^2}$ \cite{Bra}. Let us estimate $t^f_{\mathrm{ch}}$ from the relation $w_h =\sqrt{kMt^f_{\mathrm{ch}}/\eta}$, where $\eta$ is the fluid phase viscosity. This relation is obtained because fluid diffusion can be described by adding a term $(k/\eta) \partial^2p_D/\partial y^2$ to the right hand side of Eq. (\ref{eqP}). When $k$ is $10^{-21} \mathrm{m}^2$, $w_h=3 \ \mathrm{mm}$ and $3 \ \mathrm{cm}$ give $t^f_{\mathrm{ch}} \sim 8.54 \times 10 \ \mathrm{s}$ and $8.54 \times 10^3$, respectively, whereas when $k$ is $10^{-14} \ \mathrm{m}^2$, $w_h=3 \ \mathrm{mm}$ and $3 \ \mathrm{cm}$ lead to $t^f_{\mathrm{ch}} \sim 8.54 \times 10^{-6} \ \mathrm{s}$ and $8.54 \times 10^{-4}$, respectively (the values of $M$ and $\eta$ are listed in Table I).

We then consider the characteristic heat diffusion time, $t^h_{\mathrm{ch}}$, defined as $t^h_{\mathrm{ch}}=((1-\phi_t) \rho_s C_s +\phi_t \rho_f C_f) w_h^2 /\lambda$, where $\lambda$ is the thermal conductivity of the medium. This definition is reasonable because the term $\lambda \partial^2 T_D/\partial y^2$ added to the right hand side of Eq. (\ref{eqT}) describes the heat diffusion. Using the values shown in Table I and $\lambda \sim 1 \ \mathrm{J/mKs}$ \cite{Zhe}, we have $t^h_{\mathrm{ch}} \sim 2.39 \times 10 \ \mathrm{s}$ and $2.39 \times 10^3 \ \mathrm{s}$ for $w_h=3 \ \mathrm{mm}$ and $3 \ \mathrm{cm}$, respectively.

We can show the conditions for neglecting fluid and heat diffusions based on the results of $t^f_{\mathrm{ch}}$ and $t^h_{\mathrm{ch}}$. If we consider the time scale $\le$ 10 s, neglecting heat diffusion is considered to be valid because $t^h_{\mathrm{ch}} > 10$ s. Additionally, if we consider events with this time scale, we can conclude that when $k$ is near the lower limit, fluid diffusion can be neglected (i.e., the system is undrained), whereas when $k$ is near the upper limit, the diffusion effect should be taken into account (the system is drained). We can thus insist that the undrained assumption corresponds to that of lower $k$, and such low $k$ is assumed henceforth. To summarize, the approximation of neglecting fluid and heat diffusions has been found to be valid for the slip duration $\le 10 \ \mathrm{s}$ with $k$ near the lower limit. A duration lower than 10 s is characteristic time scale of earthquake duration with a moment magnitude $\le 7$, hence earthquakes with a moment magnitude $\le 7$ will be considered below. However, note that the model here developed is 1D; therefore, the moment magnitude cannot be exactly defined. This will be stated again in Sec. \ref{secApp}.

Equations (\ref{eqP}) and (\ref{eqT}) govern the temporal evolution of the fluid pressure and temperature, respectively. Equation (\ref{eqEOM}) provides a solution for the equation of motion (EOM) of the medium, with a boundary condition that difference between the applied shear stress and the frictional stress (stress drop, $\Delta \sigma$) is given by $\sigma_s^0-\sigma_{\mathrm{res}}$ \cite{Bru}. In SY14, the displacement appearing in the EOM was that of the solid phase; nonetheless, the displacements of the solid and fluid phases are exactly the same in the present system because fluid diffusion is neglected. Equation (\ref{eqSres}) provides a definition for $\sigma_{\mathrm{res}}$. Note that the normal stress acting on the fault $\sigma_n^0$ is negative because the compression stress is defined as a negative value. In particular, the first and second terms of the right hand side of Eq. (\ref{eqP}) stand for the thermal pressurization and dilatancy effects, respectively. We also assumed that $\phi_e \gg \phi_d$ and $\phi_t=\phi_e +\phi_d \sim \phi_e$ (constant) in Eqs. (\ref{eqP}-\ref{eqEOM}), which was confirmed to be reasonable from the viewpoint of laboratory experiments \cite{Mar} and numerical simulations (SY14, \cite{Suz10}).

The framework here is constructed based on SY14. In fact, researchers have shown no agreement on the mathematical treatment of the motion of the thermoporoelastic medium, even without the dilatancy effect \cite{Bio56-a, Bio56-b, Pri92, Pri93}. Despite the lack of agreement, the differences emerging in each framework do not cause any qualitative change in the system behavior because only the analytical forms of the coefficients appearing in the governing equations differ from one another. We can use the framework here as a general framework for treating ITPD.

To close the governing equation system, we also need the equation governing inelastic porosity evolution, which will be referred to as a porosity evolution law. Some forms of the law have been suggested based on many laboratory experiments, although there exists no agreement concerning its analytical form \cite{Bart, Teu, Rud, Mar, Sle}. Therefore, we should derive robust results independent of the details of the porosity evolution law. For consistency with previous studies, we only assume that $\phi_d$ is a function in terms of the slip, $u_D$, and is initially zero, i.e., $\phi_d=\phi_d (u_D)$ and $\phi_d (0)=0$. However, we require some conditions on the form of $\phi_d$ from the physical viewpoint. First, the function $\phi_d(u_D)$ is assumed to monotonically increase with increasing $u_D$ since we do not consider pore healing in an entire single slip event; this implies that $\partial \phi_d/\partial u_D$ is always nonnegative. Second, we also assume that $\displaystyle{ \lim_{u_D \to \infty} \partial \phi_d/ \partial t =0 }$, because the relation $0 \le \phi_d \le 1$ must be satisfied and $\phi_d$ must have an upper limit, which we call $\phi_{\mathrm{UL}}$ henceforth. Note that $\phi_{\mathrm{UL}}$ need not be equal to unity. From these statements, we have the porosity evolution law
\begin{equation}
\frac{\partial \phi_d}{\partial t} =\frac{\partial \phi_d}{\partial u_D} v_D, \label{eqPEL}
\end{equation}
for which the condition $\displaystyle{ \lim_{u_D \to \infty} \partial \phi_d/\partial u_D=0 }$ must be satisfied.

We can obtain the normalized governing equation system from Eqs. (\ref{eqP}-\ref{eqPEL}) as
\begin{equation}
\dot{v}=v(1-v)- \beta f(u) v, \label{eqGovG1}
\end{equation}
\begin{equation}
\dot{\phi}= f(u) v, \label{eqGovG2}
\end{equation}
where $v$ $(0 \le v \le 1)$ and $u$ are normalized slip velocity and slip, respectively, $\phi$ is normalized inelastic porosity $(0 \le \phi \le 1)$, $\beta$ is a positive constant, and $f(u)$ is a function of $u$. The overdot stands for differentiation with respect to normalized time, $\tau$. To derive Eqs. (\ref{eqGovG1}) and (\ref{eqGovG2}), note that $v_D$ and $p_D$ are linearly related (see Eqs. (\ref{eqEOM}) and (\ref{eqSres})), allowing us to rewrite $p_D$ in terms of $v_D$. The forms of $u$, $v$, $\phi$, $\beta$, $f(u)$ and $\tau$ are given by
\begin{equation}
u = \frac{((b-\phi_t) \alpha_s + \phi_t \alpha_f) M \mu_{\mathrm{slid}} }{((1-\phi_t) \rho_s C_s +\phi_t \rho_f C_f) w_h } u_D \equiv \frac{u_D}{U_{\mathrm{ref}}}, \label{eqNu}
\end{equation}
\begin{equation}
v=\frac{\mu}{2 \beta_v \sigma_s^0} v_D, \label{eqNv}
\end{equation}
\begin{equation}
\phi =\frac{\phi_d}{\phi_{\mathrm{UL}}},
\end{equation}
\begin{equation}
\beta =\frac{M \mu_{\mathrm{slid}} \phi_{\mathrm{UL}}}{\sigma_s^0}, \label{eqParBeta}
\end{equation}
\begin{equation}
f(u)=\frac{\partial}{\partial u} \phi (U_{\mathrm{ref}} u),
\end{equation}
\begin{equation}
\tau = \frac{2 \beta_v ((b-\phi_t) \alpha_s + \phi_t \alpha_f) \sigma_s^0 M \mu_{\mathrm{slid}} }{((1-\phi_t) \rho_s C_s +\phi_t \rho_f C_f) w_h \mu} t, \label{eqParTau}
\end{equation}
respectively. Based on the assumption for $\partial \phi_d/\partial u_D$, $f(u)$ is a nonnegative function. We also have the condition $\displaystyle{ \lim_{u \to \infty} f(u)=0 }$, owing to the condition $\displaystyle{ \lim_{u_D \to \infty} \partial \phi_d/\partial u_D=0 }$. Moreover, the parameter $\beta$ describes the contribution of inelastic porosity increase to the slip velocity change, because from Eqs. (\ref{eqGovG1}) and (\ref{eqGovG2}) one can obtain $\dot{v}=v(1-v)-\beta \dot{\phi}$.

In fact, the temporal evolution equation for the normalized temperature can be given by
\begin{equation}
\dot{T}=v(1-v),
\end{equation}
where $T=T_D ((b-\phi_t) \alpha_s + \phi_t \alpha_f) M \mu_{\mathrm{slid}}/\sigma_s^0$ is the normalized temperature. Nonetheless, this will not be handled in the investigation below because the governing equation system (\ref{eqGovG1}) and (\ref{eqGovG2}) is closed in terms of $v$ and $\phi$ (note that $\displaystyle{ u=\int v d\tau }$), and the temperature can be determined by calculating $\displaystyle{ T=\int v(1-v) d \tau }$.

We now describe $\phi$ in terms of $u$ for convenience during later analytical treatments. From Eq. (\ref{eqGovG2}), we have $\displaystyle{ \phi =\int f(u) du \equiv F(u) }$, so
\begin{equation}
u=F^{-1} (\phi), \label{eqRelUPhi}
\end{equation}
where $F^{-1}$ is an inverse function of $F$. Because $f(u)$ is a nonnegative function, $\displaystyle{ F(u) = \int f(u) du }$ is a monotonically increasing function in terms of $u$. Therefore, $F(u)$ clearly has its inverse function. Using Eq. (\ref{eqRelUPhi}), we have the following governing equations
\begin{equation}
\dot{v}=v(1-v)- \beta f(F^{-1}(\phi)) v, \label{eqGovG3}
\end{equation}
\begin{equation}
\dot{\phi}= f(F^{-1}(\phi)) v. \label{eqGovG4}
\end{equation}

Equations (\ref{eqGovG3}) and (\ref{eqGovG4}) form the general framework treating ITPD in the 1D model, and our analytical treatment will be based on these equations henceforth. In particular, we will derive a kind of phase transition and universal criticality emerging near the phase transition point.

\section{ATTRACTORS OBSERVED IN THE GOVERNING EQUATION SYSTEM} \label{secATT}

\subsection{Geometrically different attractors} \label{secGEO}

\begin{figure}[tbp]
\centering
\includegraphics[width=8.5cm]{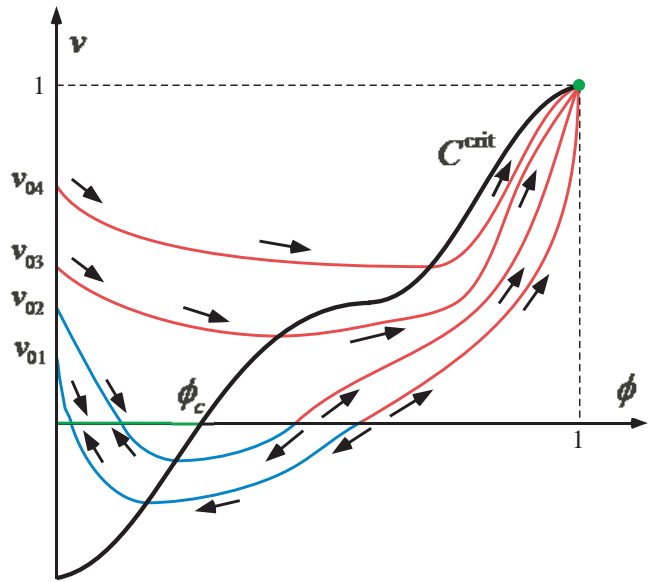}
\caption{Solution orbits. The blue and red curves are absorbed into the line and point attractors, respectively. Small arrows denote the directions of solution movement with increasing time. Points $(0, v_{01})-(0, v_{04})$ are those where the solution orbits cross the $v$-axis.}
\label{FigSO}
\end{figure}

We mathematically demonstrate that two geometrically different attractors emerge within the present framework including $v$ and $\phi$. To show this, we consider the qualitative behavior of the solution orbit in $\phi-v$ space (Fig. \ref{FigSO}). We first consider nullclines, which are obtained by the conditions $\dot{v}=0$ and $\dot{\phi}=0$. For $\dot{v}=0$, the straight line $v=0$ and the curve $v=1-\beta f(F^{-1}(\phi)) (\equiv g(\phi))$ are nullclines from Eq. (\ref{eqGovG3}). The curve $v=g(\phi)$ on the $\phi-v$ space will be referred to as $C^{\mathrm{crit}}$ henceforth. For $\dot{\phi}=0$, the straight line $v=0$ and the curve $f(F^{-1}(\phi))(=(1-g(\phi))/\beta)=0$ are nullclines. Clearly, the line $v=0$ is the nullcline for both equations. Moreover, we show here that the curve $f(F^{-1}(\phi))=0$ can be described by the straight line $\phi=1$ in $\phi-v$ space. Note that the condition $f(F^{-1}(\phi))=0$ corresponds to $F^{-1}(\phi) \to \infty$ based on the assumption $\displaystyle{ \lim_{u \to \infty} f(u)=0 }$; therefore, $\displaystyle{ \phi =\lim_{u \to \infty} F(u) }$ must be satisfied on such a nullcline. It should also be emphasized that (a) $\phi$ is a monotonically increasing continuous function in terms of $u$, and (b) $\phi$ is normalized to have a maximum value of unity. These statements suggest that the nullcline $f(F^{-1}(\phi))=0$ is given by the straight line $\displaystyle{ \phi=\lim_{u \to \infty}F(u)=1 }$. From these statements, we can also conclude that $C^{\mathrm{crit}}$ is absorbed into point $(1, 1)$. Additionally, we utilize an important condition here for $C^{\mathrm{crit}}$. We assume the relation
\begin{equation}
g(0)<0. \label{eqCon1}
\end{equation}
This assumption and the statement that $C^{\mathrm{crit}}$ is absorbed into point $(1, 1)$ allow us to conclude that $C^{\mathrm{crit}}$ crosses the $\phi$-axis in the region $0<\phi<1$, i.e., at least one positive $\phi_c$ satisfying
\begin{equation}
g(\phi_c)=0 \ \ \ \ \ (0 < \phi_c < 1) \label{eqPhic}
\end{equation}
is assumed to exist (see Fig. \ref{FigSO}). The same condition has been treated in several previous studies (SY14, \cite{Ric06}).

We require another important assumption for the solution orbit, namely that the orbits cross the $v$-axis in the region $0<v<1$; that is, we use the condition $0<v_0<1$, where $v_0$ is the value of $v$ at $\phi=0$, for the sake of physical and seismological applications in the latter part of this paper (the physically meaningful regions for $\phi$ and $v$ are $0 \le \phi \le 1$ and $0 \le v \le 1$, respectively). However, because the mathematical treatment is performed in the present subsection, the orbit can pass through the region $v<0$.

First, we assume that $g'(\phi) \ge 0$ is satisfied for $0 \le \phi \le 1$, where the prime denotes differentiation with respect to $\phi$. We also assume that if points satisfying $g'(\phi)=0$ exist, they are isolated. With these assumptions, Eq. (\ref{eqPhic}) has a single positive root (which can be a multiple one). In this case, $C^{\mathrm{crit}}$ is clearly right-upward and crosses the $\phi$-axis once (Fig. \ref{FigSO}). The sign of the gradient at the given point on the solution orbit depends on whether the point is on the upside or underside of $C^{\mathrm{crit}}$ because $dv/d \phi=0$ on $C^{\mathrm{crit}}$. If the point is on the upside of $C^{\mathrm{crit}}$, the gradient is negative, whereas if it is on the underside of $C^{\mathrm{crit}}$, the gradient is positive. The orbit becomes horizontal at the point crossing $C^{\mathrm{crit}}$. Moreover, it should be noted that the orbit is neither horizontal nor vertical at the point crossing the line $v=0$, even though the line is a nullcline. This occurs because $v=0$ is a nullcline for both equations; the condition $\dot{v}=\dot{\phi}=0$ is satisfied on the line $v=0$, enabling $dv/d \phi$ to be nonzero and finite there. From these statements and Fig. \ref{FigSO}, the orbit is found to connect the point $(0, v_0)$ with $(1, 1)$. In fact, the point $(1,0)$ is also a fixed one for Eqs. (\ref{eqGovG3}) and (\ref{eqGovG4}), and no orbit connecting $(0,v_0)$ with $(1,0)$ exists, as shown in Appendix \ref{secAA}.

Figure \ref{FigSO} also shows arrows indicating the directions of evolution of the solutions with increasing time. These arrows can easily be obtained by the relation $\dot{\phi}=f(F^{-1}(\phi)) v$, because $f(F^{-1}(\phi))$ is always positive except on the nullcline $\phi=1$, as noted in Sec. \ref{secMS}, and the direction of the solution is determined only by the sign of $v$. If $v>0$, the solution moves rightward, whereas if $v<0$, it moves leftward with increasing time. The solution orbits and the arrows shown in Fig. $\ref{FigSO}$ illustrate that we have an attractor and repeller on the $\phi$-axis; $\{ (\phi_a, 0) | \ 0 \le \phi_a \le \phi_c \}$ is an attractor, and $\{ (\phi_r, 0) | \ \phi_c \le \phi_r \le 1 \}$ is a repeller, where $\phi_a$ and $\phi_r$ are real numbers. In particular, note that $\phi_a$ and $\phi_r$ take continuous values. These non-isolated fixed points appear because the line $v=0$ is a common nullcline for both equations, and this is a noteworthy behavior of the present system. In addition, the point $(1, 1)$ is also an attractor because $C^{\mathrm{crit}}$ and all the orbits are absorbed into the point $(1, 1)$, and the solutions move rightward with increasing time where $v>0$. We can therefore categorize the attractors into two geometrically different groups, which are given by the line $ \{ (\phi_a, 0) | \ 0 \le \phi_a \le \phi_c \}$ or the point $(1,1)$ (see the green line and point in Fig. \ref{FigSO}). We will refer to the former and latter attractors as line and point attractors, respectively. Exact analytical investigation determining which attractor appears will be carried out in Sec. \ref{secPT}.

\begin{figure}[tbp]
\centering
\includegraphics[width=8.5cm]{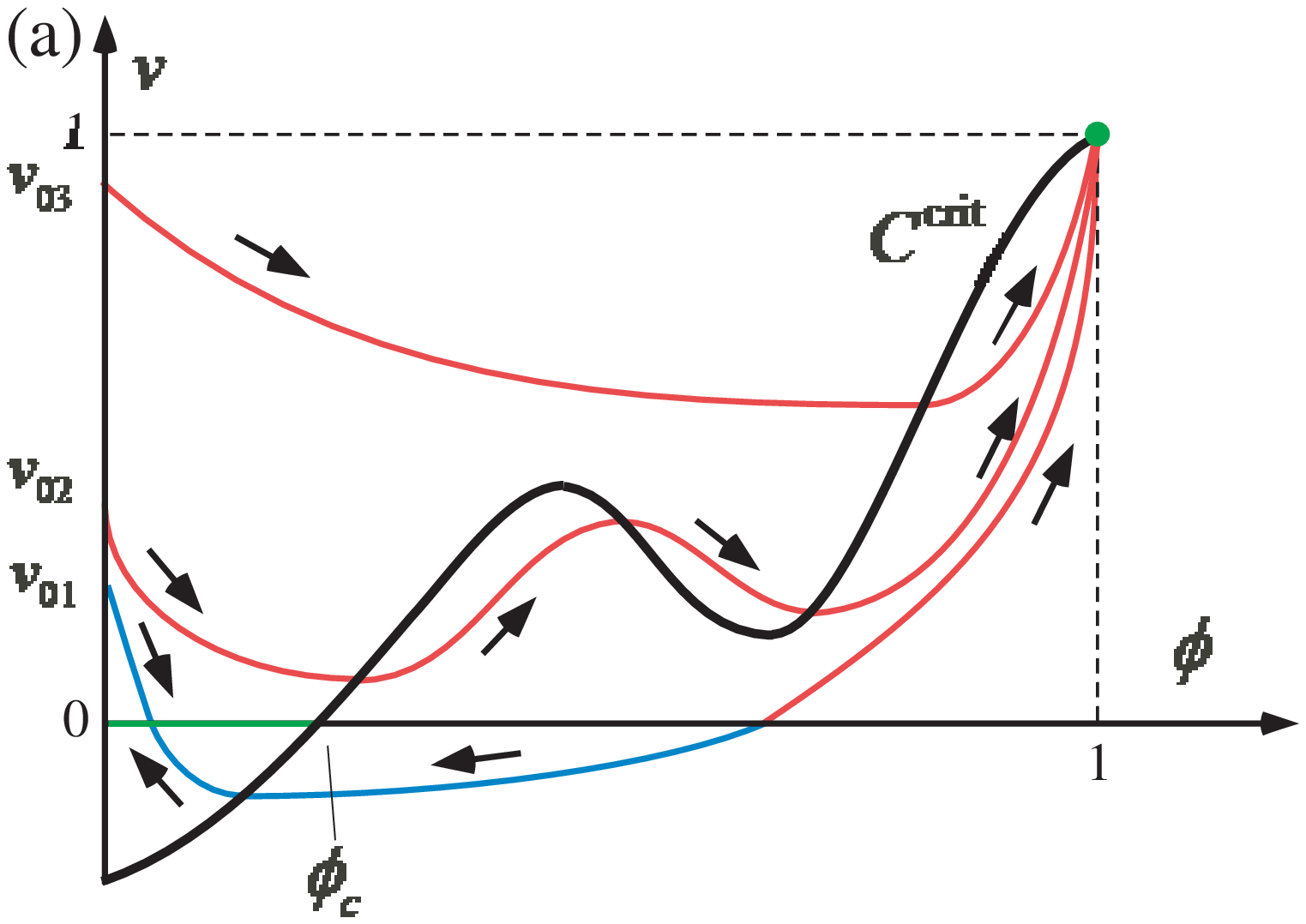}
\includegraphics[width=8.5cm]{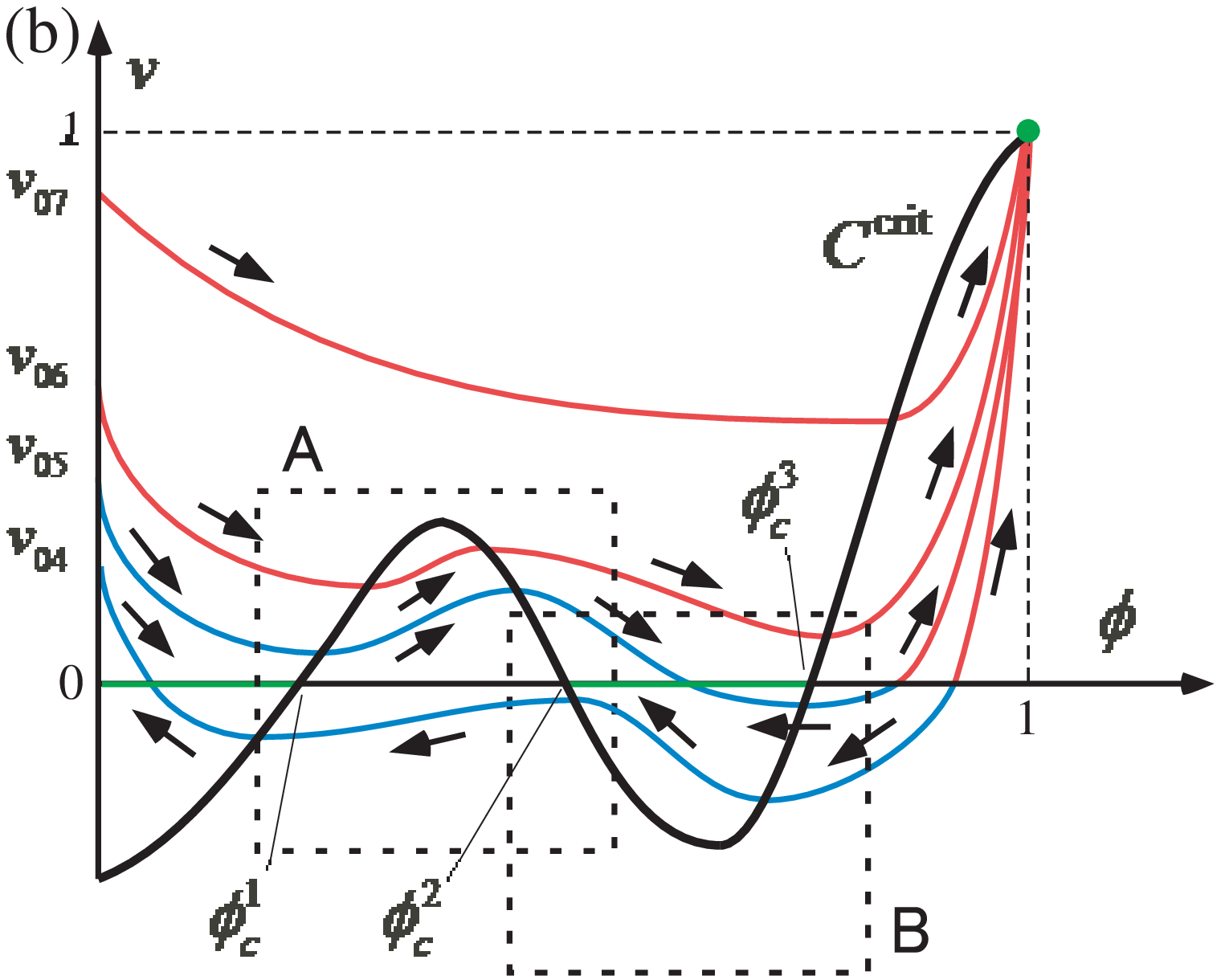}
\caption{Solution orbits. The blue and red curves are absorbed into the line and point attractors, respectively. (a) Case wherein the nullcline $C^{\mathrm{crit}}$ crosses the $\phi$-axis once. (b) Case wherein the nullcline $C^{\mathrm{crit}}$ crosses the $\phi$-axis more than once. Points $(0, v_{01})-(0, v_{07})$ are those where the solution orbits cross the $v$-axis. The dotted rectangles A and B describe the examples treated in Figs. \ref{FigSOcc}(d) and \ref{FigSOcc}(a), respectively.}
\label{FigSOmm}
\end{figure}

In Fig. \ref{FigSO}, we assumed that $g'(\phi) \ge 0$ for $0 \le \phi \le 1$ and that the curve $C^{\mathrm{crit}}$ was right-upward. Here, we allow $g'(\phi)$ to be negative at a certain value of $\phi$. Transitions from right-upward (right-downward) to right-downward (right-upward) with increasing $\phi$ (referred to as up$-$down (down$-$up) transitions henceforth) emerge with this condition. These cases correspond to the condition that $g(\phi)$ has maximal (for up$-$down transition) and minimal (for down$-$up transition) values. First, let us consider $C^{\mathrm{crit}}$ with a single up$-$down transition and a single down$-$up transition. With this condition, the case where $C^{\mathrm{crit}}$ crosses the $\phi$-axis once is shown in Fig. \ref{FigSOmm}(a). In Fig. \ref{FigSOmm}(a), although an orbit crossing $C^{\mathrm{crit}}$ three times appears, the attractors are the same as those observed in Fig. \ref{FigSO}. Additionally, the case where $C^{\mathrm{crit}}$ crosses the $\phi$-axis more than once is shown in Fig. \ref{FigSOmm}(b). When there exist two or more $\phi_c$, we refer to the $m$th smallest one as $\phi_c^m$, where $m$ is a positive integer. When $\phi_c$ is a multiple root, it is regarded as a single $\phi_c^m$. In this case, the area where $v < g(\phi)$ (i.e., the underside of $C^{\mathrm{crit}}$) cannot include attractors for the following reasons. The solution must move away from the $\phi$-axis there because (a) $dv/d \phi >0$ is satisfied and (b) the solution describing infinitesimal perturbation from the $\phi$-axis to the positive (negative) $v$ direction moves rightward (leftward) with increasing time. Therefore, there exist two line attractors on the $\phi$-axis, and the point $(1,1)$ is also an attractor. We can insist that even if $C^{\mathrm{crit}}$ is not monotonically rising with increasing $\phi$, two geometrically different attractors emerge and only the number of line attractors changes.

\begin{figure*}[tbp]
\centering
\begin{tabular}{ccc}
\begin{minipage}[t]{5.cm}
\includegraphics[width=5.cm]{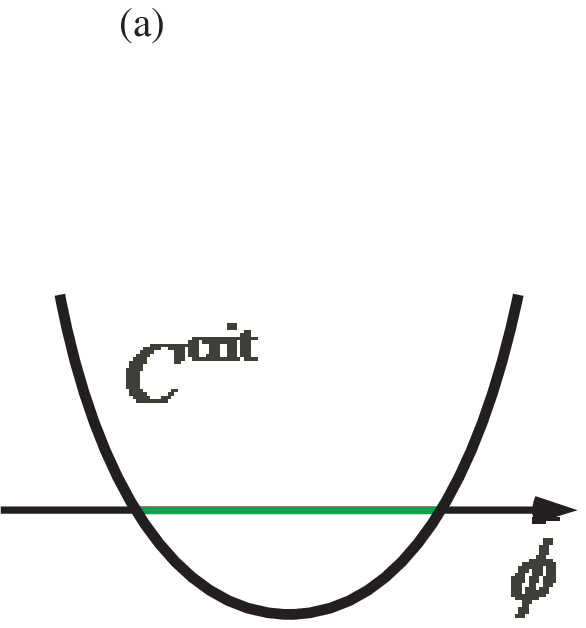}
\end{minipage} &
\begin{minipage}[t]{5.cm}
\includegraphics[width=5.cm]{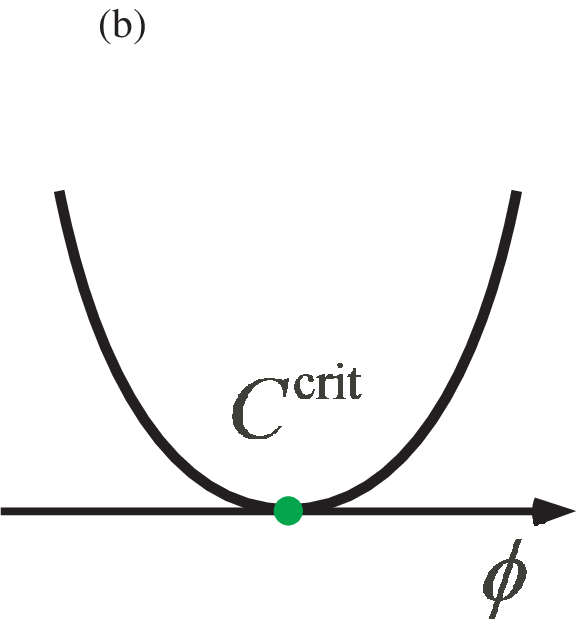}
\end{minipage} &
\begin{minipage}[t]{5.cm}
\includegraphics[width=5.cm]{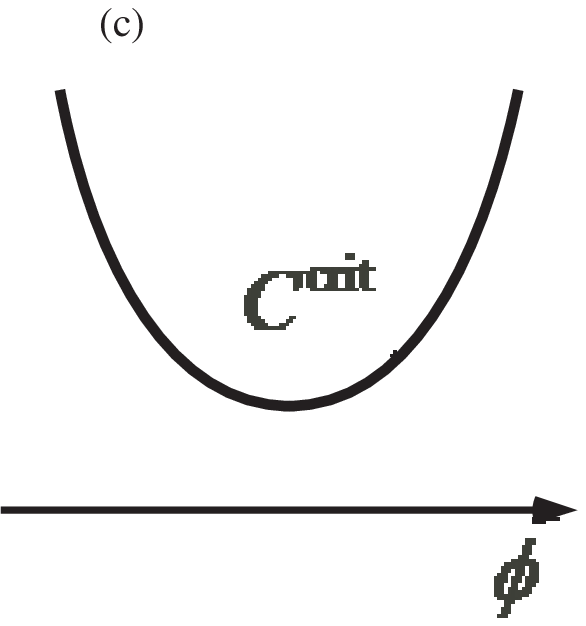}
\end{minipage}
\end{tabular} \\
\begin{tabular}{ccc}
\begin{minipage}[t]{5.cm}
\vspace{-5.3cm}
\includegraphics[width=5.cm]{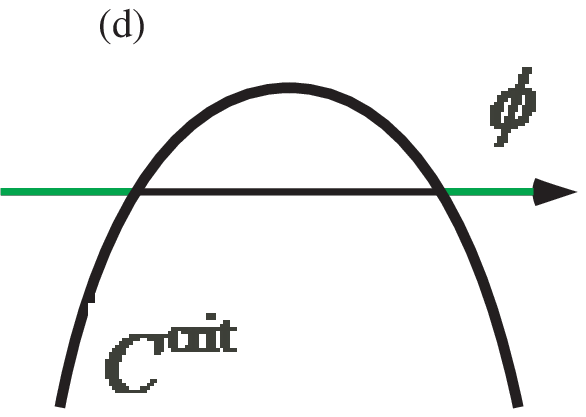}
\end{minipage} &
\begin{minipage}[t]{5.cm}
\vspace{-5.3cm}
\includegraphics[width=5.cm]{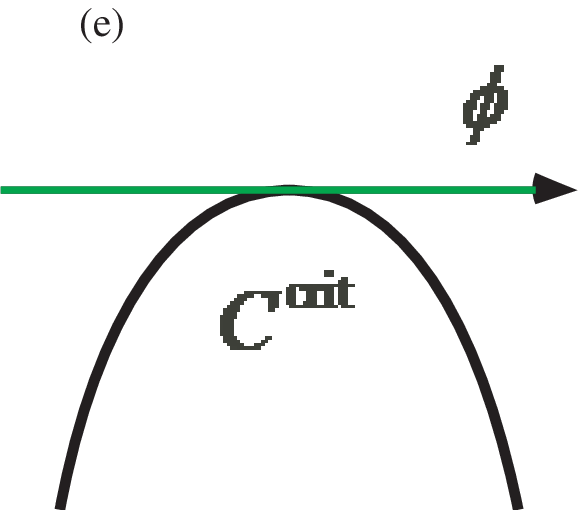}
\end{minipage} &
\begin{minipage}[t]{5.cm}
\includegraphics[width=5.cm]{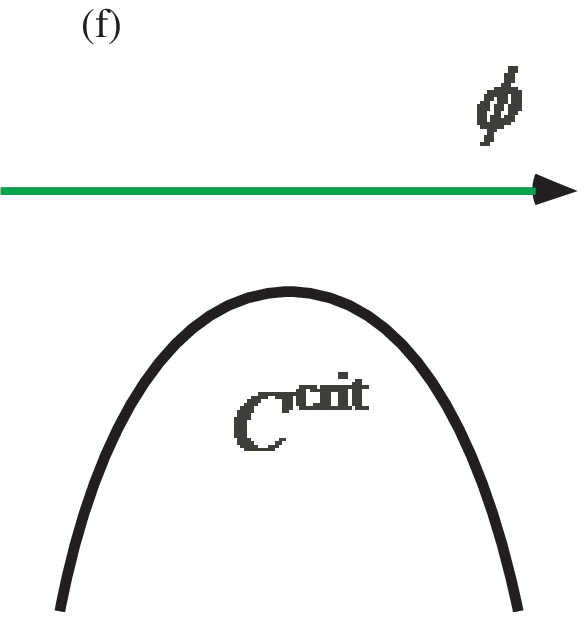}
\end{minipage}
\end{tabular}
\caption{Down$-$up and up$-$down transitions of $C^{\mathrm{crit}}$. Green lines describe line attractors, and a point such that indicated in green color represents a converged one. (a) The down$-$up transition emerges on the underside of the $\phi$-axis, and $C^{\mathrm{crit}}$ crosses the $\phi$-axis. (b) The $\phi$-axis is tangential to the down$-$up transition. (c) The down$-$up transition occurs above the $\phi$-axis. (d) The up$-$down transition emerges on the upside of the $\phi$-axis, and $C^{\mathrm{crit}}$ crosses the $\phi$-axis. (e) The $\phi$-axis is tangential to the up$-$down transition. (f) The up$-$down transition occurs on the underside of the $\phi$-axis.}
\label{FigSOcc}
\end{figure*}

We should investigate the behavior of line attractors near transition points in detail. Let us first consider the case where $C^{\mathrm{crit}}$ has a down$-$up transition. If the transition occurs under the $\phi$-axis, $C^{\mathrm{crit}}$ crosses the $\phi$-axis and the line attractor appears (Fig. \ref{FigSOcc}(a)). However, if the $\phi$-axis becomes tangent to $C^{\mathrm{crit}}$, the attractor converges to a point (Fig. \ref{FigSOcc}(b)). This converged line attractor is a special case for the line attractor, and we do not refer to it as a point attractor. If the down$-$up transition occurs above the $\phi$-axis, the attractor vanishes (Fig. \ref{FigSOcc}(c)). We now treat the case in which $C^{\mathrm{crit}}$ has an up$-$down transition. If such a transition occurs above the $\phi$-axis, $C^{\mathrm{crit}}$ crosses the $\phi$-axis and two line attractors divided by $C^{\mathrm{crit}}$ emerge in the vicinity of the transition point (Fig. \ref{FigSOcc}(d)). However, if the $\phi$-axis becomes tangential to $C^{\mathrm{crit}}$, the line attractors coalesce and turn into a single line attractor (Fig. \ref{FigSOcc}(e)). If the up$-$down transition occurs under the $\phi$-axis, a single line attractor will be observed (Fig. \ref{FigSOcc}(f)).

\begin{figure*}[tbp]
\centering
\begin{tabular}{cc}
\begin{minipage}[t]{5cm}
\includegraphics[width=5.cm]{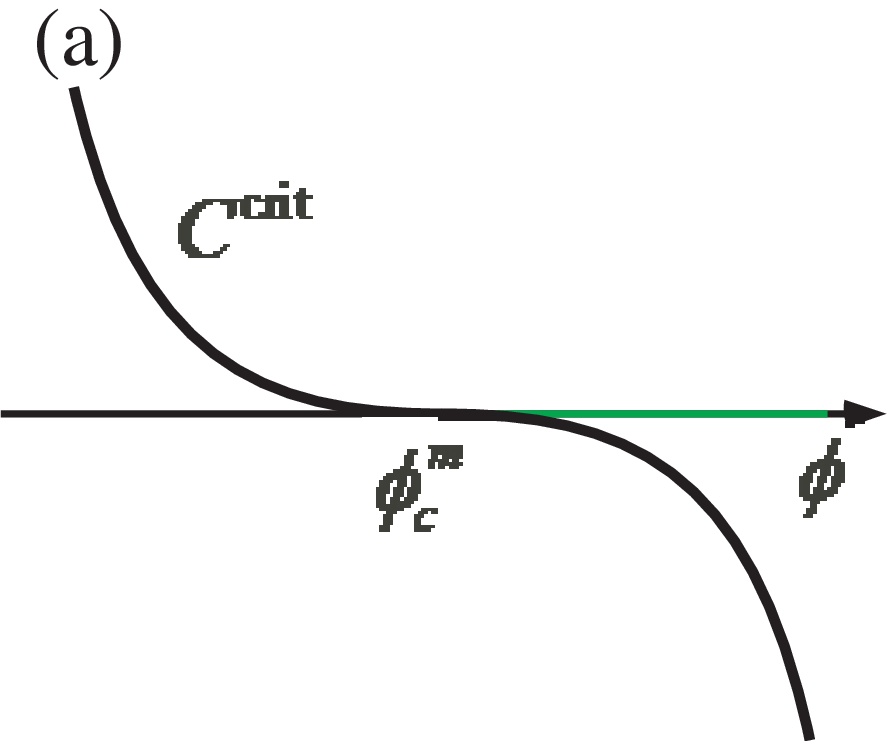}
\end{minipage} &
\begin{minipage}[t]{5cm}
\includegraphics[width=5.cm]{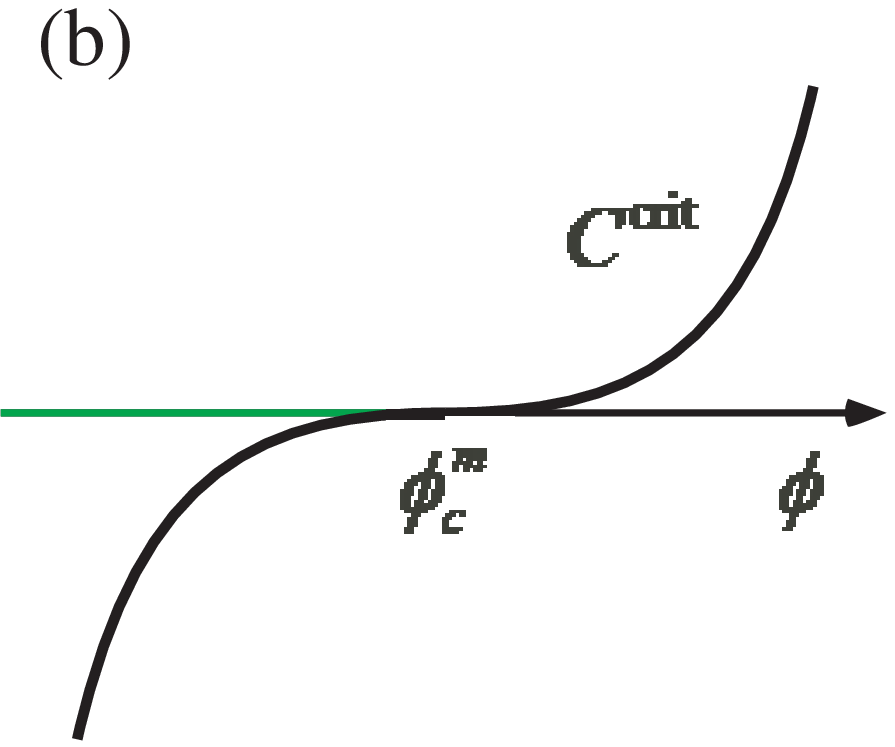}
\end{minipage}
\end{tabular}
\begin{tabular}{cc}
\begin{minipage}[t]{5cm}
\includegraphics[width=5.cm]{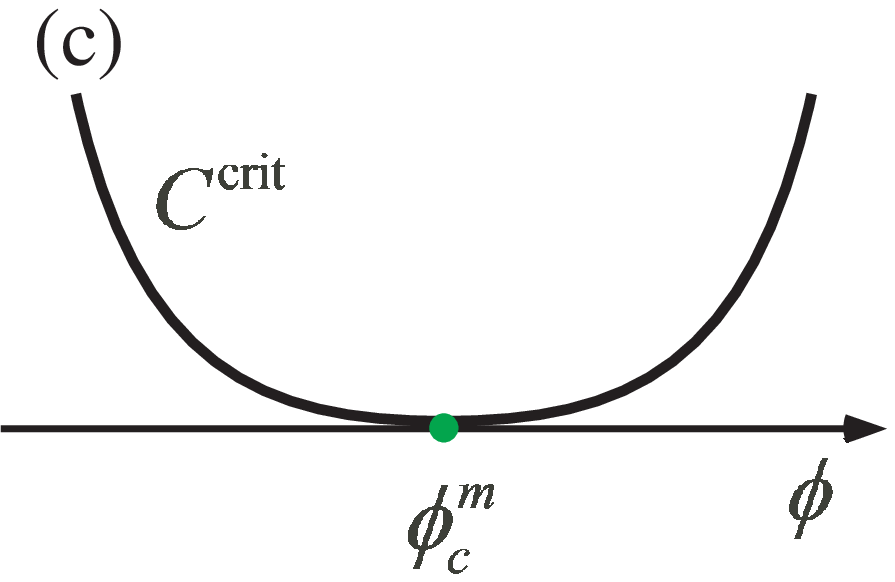}
\end{minipage} &
\begin{minipage}[t]{5cm}
\vspace{-3.25cm}
\includegraphics[width=5.cm]{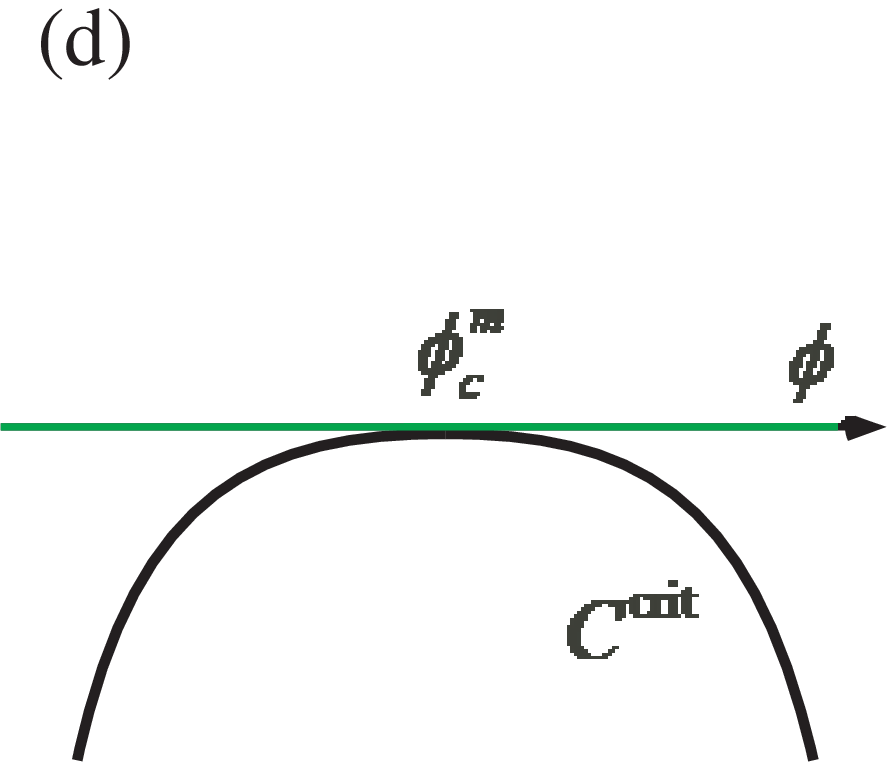}
\end{minipage}
\end{tabular}
\caption{Cases wherein the $\phi$-axis is tangential to $C^{\mathrm{crit}}$.  Green lines describe line attractors, and a point such that indicated in green color represents a converged one. The values $n$ and $g^{(n)}(\phi_c^m) (\neq 0)$ are (a) odd and negative, (b) odd and positive, (c) even and positive, and (d) even and negative. The points $(\phi_c^m, 0)$ are (a) the left end point of the line attractor, (b) the right end point of the line attractor, (c) the converged line attractor, and (d) neither the left nor the right end points of the line attractor.}
\label{FigSOtan}
\end{figure*}

\begin{figure}[tbp]
\centering
\includegraphics[width=8.5cm]{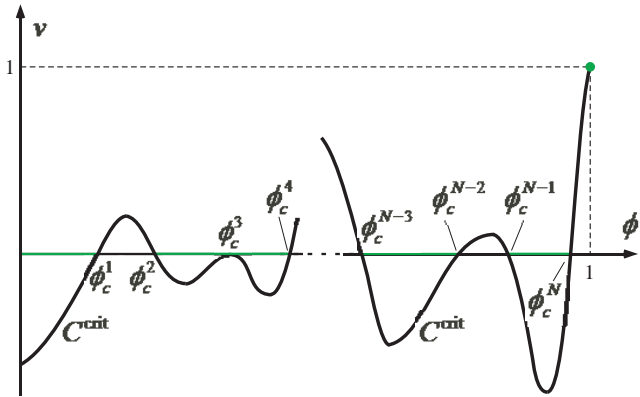}
\caption{General form of $C^{\mathrm{crit}}$. The $m$th smallest positive solution of $g(\phi)=0$ is described as $\phi_c^m$. The equation is assumed to have $N$ different solutions, where $N$ is a natural number. In this example, $\phi_c^3$ is a multiple root of $g(\phi)=0$.}
\label{FigSOgen}
\end{figure}

\begin{figure}[htbp]
\centering
\includegraphics[width=8.5cm]{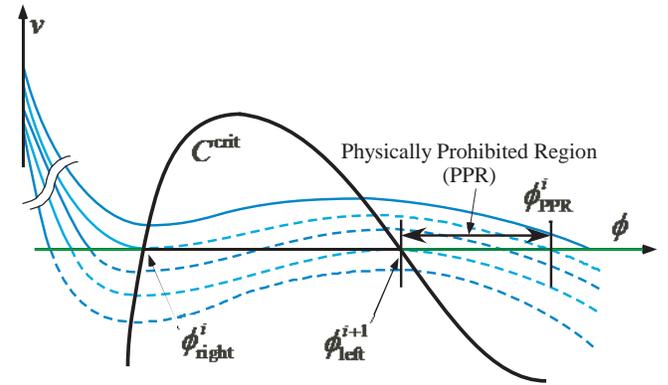}
\caption{Physically prohibited region. The light blue curves cross $(\phi_c^m, 0)$. The solid parts of the orbits are physically meaningful, whereas the dotted parts represent intervals of the mathematical solution which do not have a real physical meaning. The physically meaningful orbits cannot approach the region labeled as PPR, even though the region is on the line attractor.}
\label{FigPPR}
\end{figure}

\begin{figure}[tbp]
\centering
\includegraphics[width=8.5cm]{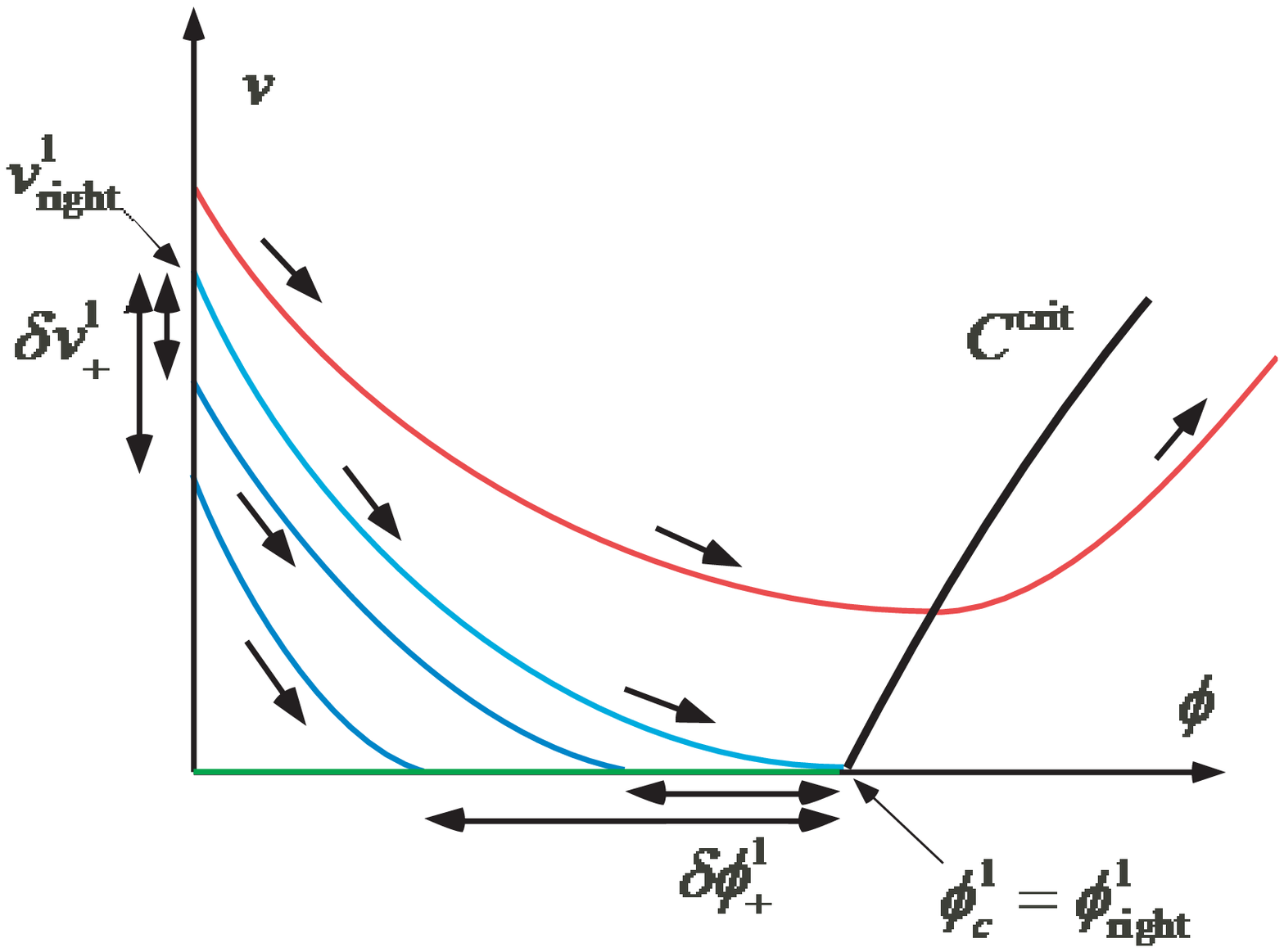}
\caption{Solution orbits whose configuration is an enlarged version of Fig. \ref{FigSO} in the vicinity of $(\phi_{\mathrm{right}}^1, 0)$. The light blue curve represents the critical manifold. The blue curves illustrate the solution orbit absorbed into the first line attractor. The red curve indicates the solution orbit absorbed into the point attractor. Short black arrows show the direction of solution movement with increasing time. Two variables, $\delta v_+^1$ and $\delta \phi_+^1$, for two orbits are also shown. See details in the text.}
\label{FigSOen}
\end{figure}

We can summarize the behaviors of the line attractors in terms of $g$ and its derivatives. Let $\forall j, \ g^{(j)}(\phi_c^m)=0$ and $g^{(n)}(\phi_c^m) \neq 0$, where $g^{(j)}$ stands for the $j$th derivative of $g$ with respect to $\phi$, $n$ is a positive integer and $j$ is a nonnegative integer satisfying $0 \le j \le n-1$ (note that $g^{(0)}=g$). If $n$ is odd and $g^{(n)}(\phi_c^m)<0$, the point $(\phi_c^m, 0)$ is the left end point of a line attractor, as $\phi_c^2$ in Fig. \ref{FigSOmm}(b) and $\phi_c^m$ in Fig. \ref{FigSOtan}(a). If $n$ is odd and $g^{(n)}(\phi_c^m)>0$, the point $(\phi_c^m, 0)$ is the right end point of a line attractor, as $\phi_c^1$ and $\phi_c^3$ in Fig. \ref{FigSOmm}(b) and $\phi_c^m$ in Fig. \ref{FigSOtan}(b). If $n$ is even and $g^{(n)}(\phi_c^m)>0$, the converged line attractor is generated at $(\phi_c^m, 0)$ as shown in Figs. \ref{FigSOcc}(b) and \ref{FigSOtan}(c). If $n$ is even and $g^{(n)}(\phi_c^m)<0$, the coalesced line attractor is observed as in Figs. \ref{FigSOcc}(e) and \ref{FigSOtan}(d). The general form of the line attractors is shown in Fig. \ref{FigSOgen}. The value of $\phi_c^m$ at the left (right) end of the $i$th line attractor from left will be referred to as $\phi_{\mathrm{left}}^i$ ($\phi_{\mathrm{right}}^i$) henceforth (by definition, $\phi_{\mathrm{left}}^1=0$), where $i$ is a positive integer. When the converged line attractor emerges at $(\phi_c^m, 0)$, the point is regarded as both the left and the right end points, whereas when the line attractors coalesce at $(\phi_c^m, 0)$, the point is neither the left nor the right end point. Finally, we can also emphasize that the emergence of two geometrically different attractors does not depend on the details of $g(\phi)(=1-\beta f(F^{-1}(\phi)))$, i.e., the porosity evolution law. Only the number of line attractors varies owing to this detail.

\subsection{Important concepts from the physical viewpoint} \label{secPI}

Here, we introduce some physically meaningful concepts associated with the current mathematical treatment. First, we introduce ``phases'' of the attractors for later discussion about phase transition. The line and point attractors physically correspond to $\displaystyle{\lim_{\tau \to \infty} v=0 }$ and $1$, respectively. Hence, we can refer to the former and latter cases as the cessation and high-speed phases, respectively, from the physical viewpoint. The physical elementary process realizing these phases will be explained in Sec. \ref{secApp}. 

We now clarify the physically meaningful solution orbits. For the cases shown in Fig. \ref{FigSO}, the orbit starts on the $v$-axis ($\phi=0$) physically. Within the orbits beginning with $v_0=v_{01}$ or $v_{02}$ in Fig. \ref{FigSO}, only the parts before absorption into the green line are physical, and the other parts are mathematical and unphysical. Additionally, for the cases shown in Fig. \ref{FigSOmm}(b), the solution orbits crossing the green lines are physical before absorption, which is the same as in Fig. \ref{FigSO}. However, a physically prohibited region (PPR) can emerge near $\phi_{\mathrm{left}}^{i+1}$ (Fig. \ref{FigPPR}). We define the point $(\phi_{\mathrm{PPR}}^i, 0)$ as being where the orbit crossing the point $(\phi_{\mathrm{right}}^i, 0)$ again crosses the $\phi$-axis. If such an orbit is absorbed into the point attractor, we define $\phi_{\mathrm{PPR}}^i=1$. The solutions cannot be absorbed into the region $\{ (\phi, 0)| \ \phi_{\mathrm{left}}^{i+1} \le \phi \le \phi_{\mathrm{PPR}}^i \}$ with a physically meaningful initial condition and infinitely long time, and this region will be referred to as the PPR. The left end points of the line attractors other than the origin $(\phi_{\mathrm{left}}^1, 0)$ must be accompanied by the PPR. If $\phi_{\mathrm{right}}^{i+1} < \phi_{\mathrm{PPR}}^i$, the solutions cannot be absorbed into the $i+1$th line attractor. To summarize, although all green lines in Fig. \ref{FigSOgen} actually represent line attractors in a mathematical sense, not all points on the attractors describe physically meaningful solutions.

\section{PHASE TRANSITION AND UNIVERSALITY} \label{secPT}

From the investigation above, we can conclude that a system including ITPD generates a phase transition between the cessation and the high-speed phases. We now show that universal (scale-independent) behavior emerges in the vicinity of the phase transition point in the present model by considering the perturbation in $v_0$, which is similar to other phase transitions, e.g., susceptibility in phase transition of the second kind, and the spanning probability of percolation \cite{Flo, Sta}. Such universality has many implications for the behavior of the final slip amount, which plays an important role in our understanding of the dynamic earthquake slip process, as shown  in Sec. \ref{secApp}. 

To derive the universality, see the area near the origin in Fig. \ref{FigSO}, which is enlarged in Fig. \ref{FigSOen}. In this case, $C^{\mathrm{crit}}$ is assumed to cross the $\phi$-axis once at $(\phi_c^1, 0)$ and $g'(\phi_c^1)>0$, and we have the single solution $\phi_c=\phi_c^1=\phi_{\mathrm{right}}^1$ for Eq. (\ref{eqPhic}). Note that we have a manifold dividing the point attractor and the line attractor (i.e., the basin boundary), which is drawn by the light blue curve in Fig. \ref{FigSOen}. We refer to the manifold as the critical manifold, and it begins at $(0, v_{\mathrm{right}}^1)$ and ends at $(\phi_{\mathrm{right}}^1, 0)$, where $(0, v_{\mathrm{right}}^i)$ is a point where the manifold crossing $(\phi_{\mathrm{right}}^i, 0)$ crosses the $v$-axis. If $v_0>v_{\mathrm{right}}^1$, the point attractor emerges, whereas if $v_0<v_{\mathrm{right}}^1$, the line attractor is realized. Let us consider here the manifolds in the neighborhood of the critical manifold. We assume that these manifolds are absorbed into the line attractor, such that the assumption $v_0<v_{\mathrm{right}}^1$ is used here. In the $\phi-v$ space, if the initiation point of the manifold is given by $(0, v_{\mathrm{right}}^1-\delta v_+^1)$, its ending point is expected to be described by $(\phi_{\mathrm{right}}^1 -\delta \phi_+^1, 0)$, where $\delta v_+^1$ and $\delta \phi_+^1$ are positive amounts that satisfy $\delta v_+^1 \ll 1$ and $\delta \phi_+^1 \ll 1$. We will show that $\delta \phi_+^1$ and $\delta v_+^1$ are related via a simple power law,
\begin{equation}
\delta \phi_+^1 \propto (\delta v_+^1)^{\alpha}, \label{eqPgen}
\end{equation}
and obtain the value of the critical exponent $\alpha$ in the following part of this section. Moreover, we will show that the details of the porosity evolution law do not affect the critical exponent.

\subsection{Derivation of the power law} \label{secDPL}

To show the power law (\ref{eqPgen}), note that Eqs. (\ref{eqGovG1}) and (\ref{eqGovG2}) yield
\begin{equation}
\frac{d v}{d \phi} =\frac{1-v-\beta f(F^{-1}(\phi))}{f(F^{-1}(\phi))}=\beta \frac{1-v}{1-g(\phi)} -\beta. \label{eqGovl}
\end{equation}
From Eq. (\ref{eqGovl}), we have the solution for $v$ based on the method of variation of constants in terms of $\phi$, which leads to
\begin{equation}
v = -\beta e^{-\beta A(\phi)} (B(\phi)-B(0)) +e^{-\beta (A(\phi)-A(0))} (v_0-1)+1, \label{eqSolPhi}
\end{equation}
where $\displaystyle{ A(\phi) \equiv \int^{\phi} d \phi^{\ast} /(1-g(\phi^{\ast})) }$ and $\displaystyle{ B(\phi) \equiv \int^{\phi} e^{\beta A(\phi^{\ast})} d \phi^{\ast} }$. The initial condition $v|_{\tau=0}=v_0$ at $\phi|_{\tau=0}=0$ is used. Equation ($\ref{eqSolPhi}$) gives the solution orbit in $\phi-v$ space.

We then obtain the value of $\phi_{\mathrm{right}}^1$ from Eq. (\ref{eqPhic}). This equation reads as
\begin{equation}
\phi_{\mathrm{right}}^1=g^{-1} (0), \label{eqSolPhic}
\end{equation}
where $g^{-1}$ is the inverse function of $g$. The value $g^{-1}(0)$ can be uniquely defined within the current framework. Note that the character $\phi_{\mathrm{right}}^1$ remains in the analytical treatment below to simplify the description because using $g^{-1}(0)$ makes the representation complex. If we do not use $\phi_{\mathrm{right}}^1$ explicitly, we should employ solution (\ref{eqSolPhic}).

Next, we obtain $v_{\mathrm{right}}^1$ from Eq. (\ref{eqSolPhi}). This value can be derived from the condition $v=0$ with $v_0=v_{\mathrm{right}}^1$ and $\phi=\phi_{\mathrm{right}}^1$ in Eq. (\ref{eqSolPhi}):
\begin{eqnarray}
&-&\beta e^{-\beta A(\phi_{\mathrm{right}}^1)} (B(\phi_{\mathrm{right}}^1)-B(0)) \nonumber \\
&+&e^{-\beta (A(\phi_{\mathrm{right}}^1)-A(0))} (v_{\mathrm{right}}^1-1)+1=0,
\end{eqnarray}
which reads as
\begin{equation}
v_{\mathrm{right}}^1=\beta e^{-\beta A(0)} (B(\phi_{\mathrm{right}}^1)-B(0))-e^{\beta (A(\phi_{\mathrm{right}}^1)-A(0))} +1. \label{eqvc}
\end{equation}
For the existence of a line attractor, the condition $v_{\mathrm{right}}^1>0$ must be satisfied, which is guaranteed if Eq. (\ref{eqCon1}) is satisfied, as shown in Appendix \ref{secAc}.

We thus consider the behavior of the solution orbit near the critical manifold and absorbed into the line attractor. We assume $v_0=v_{\mathrm{right}}^1 -\delta v_+^1$ and $\phi_{\infty}=\phi_{\mathrm{right}}^1 -\delta \phi_+^1$, where $\displaystyle{ \phi_{\infty} \equiv \lim_{\tau \to \infty} \phi }$ is the final inelastic porosity. From Eqs. (\ref{eqSolPhi}) and (\ref{eqvc}), we have
\begin{eqnarray}
&-&\beta e^{-\beta A(\phi_{\mathrm{right}}^1-\delta \phi_+^1)} (B(\phi_{\mathrm{right}}^1-\delta \phi_+^1)-B(0)) \nonumber \\
&+&e^{-\beta (A(\phi_{\mathrm{right}}^1-\delta \phi_+^1)-A(0))} ( \beta e^{-\beta A(0)} (B(\phi_{\mathrm{right}}^1)-B(0)) \nonumber \\
&-&e^{\beta (A(\phi_{\mathrm{right}}^1)-A(0))}-\delta v_+^1) +1=0. \label{eqEx1}
\end{eqnarray}
Note here that the expansion of $B(\phi_{\mathrm{right}}^1-\delta \phi_+^1)$ can be given by
\begin{eqnarray}
B(\phi_{\mathrm{right}}^1-\delta \phi_+^1) = &B&(\phi_{\mathrm{right}}^1) \nonumber \\
&-&e^{\beta A(\phi_{\mathrm{right}}^1)} \delta \phi_+^1 +\frac{\beta}{2} e^{\beta A(\phi_{\mathrm{right}}^1)} (\delta \phi_+^1)^2 \nonumber \\
&+&O((\delta \phi_+^1)^3), \label{eqEx2}
\end{eqnarray}
because we have the relations
\begin{equation}
\frac{dB}{d \phi} \Big|_{\phi=\phi_{\mathrm{right}}^1} =e^{\beta A(\phi_{\mathrm{right}}^1)},
\end{equation}
and
\begin{eqnarray}
\frac{d^2 B}{d \phi^2} \Big|_{\phi=\phi_{\mathrm{right}}^1} &=& \frac{d}{d \phi} e^{\beta A(\phi)} \Big|_{\phi=\phi_{\mathrm{right}}^1} \nonumber \\
&=&\beta \frac{e^{\beta A(\phi)}}{1-g(\phi)} \Big|_{\phi=\phi_{\mathrm{right}}^1}=\beta e^{\beta A(\phi_{\mathrm{right}}^1)} \label{eqA2}
\end{eqnarray}
(see also Eq. (\ref{eqSolPhic})). Equations (\ref{eqEx1}) and (\ref{eqEx2}) lead to
\begin{eqnarray}
\beta \delta \phi_+^1 -\frac{\beta^2}{2} (\delta \phi_+^1)^2 -1 -e^{-\beta( A(\phi_{\mathrm{right}}^1)-A(0))} \delta v_+^1 \nonumber \\
+e^{\beta( A(\phi_{\mathrm{right}}^1-\delta \phi_+^1)-A(\phi_{\mathrm{right}}^1))} +O((\delta \phi_+^1)^3) =0. \label{eqEx3}
\end{eqnarray}
We have multiplied both sides of Eq. (\ref{eqEx1}) by $\exp(\beta(A(\phi_{\mathrm{right}}^1-\delta \phi_+^1)-A(\phi_{\mathrm{right}}^1)))$. Additionally, we have another important expansion in terms of $\delta \phi_+^1$:
\begin{eqnarray}
&A&(\phi_{\mathrm{right}}^1-\delta \phi_+^1)-A(\phi_{\mathrm{right}}^1) \nonumber \\
&=&-\frac{d A(\phi)}{d \phi} \Big|_{\phi=\phi_{\mathrm{right}}^1} \delta \phi_+^1 +\frac{1}{2} \frac{d^2 A(\phi)}{d \phi^2} \Big|_{\phi=\phi_{\mathrm{right}}^1} (\delta \phi_+^1)^2 \nonumber \\
&+&O((\delta \phi_+^1)^3) \nonumber \\
&=&-\frac{\delta \phi_+^1}{1-g(\phi_{\mathrm{right}}^1)} +\frac{g'(\phi_{\mathrm{right}}^1)}{2 (1-g(\phi_{\mathrm{right}}^1))^2} (\delta \phi_+^1)^2 +O((\delta \phi_+^1)^3) \nonumber \\
&=&- \delta \phi_+^1 +\frac{g'(\phi_{\mathrm{right}}^1)}{2} (\delta \phi_+^1)^2 +O((\delta \phi_+^1)^3). \label{eqEx4}
\end{eqnarray}
Using Eqs. (\ref{eqEx3}) and (\ref{eqEx4}) and expanding the exponential function in Eq. (\ref{eqEx3}), we find that the terms of orders $(\delta \phi_+^1)^0$ and $(\delta \phi_+^1)^1$ vanish, leaving us with
\begin{equation}
\frac{\beta g'(\phi_{\mathrm{right}}^1)}{2} (\delta \phi_+^1)^2 -e^{-\beta( A(\phi_{\mathrm{right}}^1)-A(0))} \delta v_+^1 +O((\delta \phi_+^1)^3) =0.
\end{equation}
Neglecting the terms of order $(\delta \phi_+^1)^3$ and higher, we obtain a simple power law between $\delta \phi_+^1$ and $\delta v_+^1$:
\begin{equation}
\delta \phi_+^1 = (\delta v_+^1)^{1/2} \sqrt{ \frac{2 e^{\beta( A(0)-A(\phi_{\mathrm{right}}^1))}}{\beta g' (\phi_{\mathrm{right}}^1)} }. \label{eqP2}
\end{equation}
Note that we have assumed $g'(\phi_{\mathrm{right}}^1) >0$. The character $\phi_{\mathrm{right}}^1$ vanishes by using Eq. (\ref{eqSolPhic}), which leads to a power law relation described in terms of $\beta$ and $g(\phi)$. We should emphasize that the power 1/2 is universal and does not depend on either $\beta$ or the details of $g(\phi)$.

We also have the other right end points, $\phi_{\mathrm{right}}^{i_1}$ $(i_1 \ge 2)$. Furthermore, we should treat the case $g'(\phi_c^m) \le 0$. However, as noted in Sec. \ref{secPI}, the end points other than $\phi_{\mathrm{right}}^1$ and $\phi_{\mathrm{left}}^1$ can be in the PPR depending on the form of $C^{\mathrm{crit}}$, and physically natural solutions may not approach their vicinity. In addition, the possibility that the $\phi$-axis actually becomes tangential to $C^{\mathrm{crit}}$ (i.e., $g'(\phi_c^m)=0$) in natural fault is considered to be negligibly low. We will consider the region only near $(\phi_{\mathrm{right}}^1, 0)$ from the physical and seismological viewpoints in Sec. \ref{secApp}, and mathematical discussions about the other $\phi_c^m$ will be performed in Sec. \ref{secDisCon}.

\subsection{Other important suggestions about the phase transition and the power law} \label{secPLS}

We have another important conclusion about the parameters governing the phase transition based on the result obtained in this section. Note that the sign of $v_0-v_{\mathrm{right}}^1$ is concluded to be important for the phase transition. If it is positive, the high-speed phase emerges, whereas if it is negative, the cessation phase is realized. Based on Eq. (\ref{eqvc}), we have the relation
\begin{eqnarray}
v_0-v_{\mathrm{right}}^1=v_0 &-& \beta e^{-\beta A(0)} (B(\phi_{\mathrm{right}}^1)-B(0)) \nonumber \\
&+& e^{\beta (A(\phi_{\mathrm{right}}^1)-A(0))} -1, \label{eqGS}
\end{eqnarray}
which includes the parameter $\beta$ and the functions $A(\phi_{\mathrm{right}}^1)$ and $B(\phi_{\mathrm{right}}^1)$. These functions can be described by $g(\phi)$ by definition, and $\phi_{\mathrm{right}}^1$ can be written in terms of $g^{-1}$ (see Eq. (\ref{eqSolPhic})). The parameters $\beta$ and $v_0$ as well as the function $g(\phi)$ are found to govern the phase transition, and the governing function is given by the right hand side of Eq. (\ref{eqGS}).

We can also obtain a relation similar to Eq. (\ref{eqP2}) by considering the value of $\displaystyle{ u=\int v d \tau }$. We consider the region near $(\phi_{\mathrm{right}}^1,0)$ here. Let us assume that $u_{\infty}=u_{\mathrm{right}}^1-\delta u_+^1$ for the cessation phase, where $\displaystyle{ u_{\infty} \equiv \lim_{\tau \to 0} u }$ is the final slip amount, $u_{\mathrm{right}}^1=u_{\infty}|_{v_0=v_{\mathrm{right}}^1}$ and $\delta u_+^1$ is the positive amount satisfying $\delta u_+^1 \ll 1$. The relation $\delta \phi_+^1 = (1-g(\phi_{\mathrm{right}}^1)) \delta u_+^1/\beta =\delta u_+^1 /\beta$ (derived from Eqs. (\ref{eqGovG4}) and (\ref{eqSolPhic})) and Eq. (\ref{eqP2}) give the simple power law between $\delta u_+^1$ and $\delta v_+^1$:
\begin{equation}
\delta u_+^1 = (\delta v_+^1)^{1/2} \sqrt{ \frac{2 \beta e^{\beta(A(0)-A(\phi_{\mathrm{right}}^1))}}{g' (\phi_{\mathrm{right}}^1)} }. \label{eqP3}
\end{equation}
Thus, we have a power law with the same critical exponent 1/2 as observed in Eq. (\ref{eqP2}).

Equations (\ref{eqP2}) and (\ref{eqP3}) derived from the mathematical viewpoint imply that, on the phase space, the values on two line segments, one for the initial values and the other for the final values, are related via a simple power law from a physical viewpoint. Physical and seismological implications associated with this statement are given in Sec. \ref{secApp}. Note that we only consider the region in the vicinity of the point $(\phi_{\mathrm{right}}^1,0)$, and write $\phi_c$, $v_c$, $u_c$, $\delta v$, and $\delta u$ instead of  $\phi_{\mathrm{right}}^1$, $v_{\mathrm{right}}^1$, $u_{\mathrm{right}}^1$, $\delta v_{\mathrm{right}}^1$, and $\delta u_{\mathrm{right}}^1$, respectively, below.

\section{APPLICATION TO NATURAL FAULTS} \label{secApp}

Dynamic earthquake slip processes show phase transitions by considering $\phi_{\infty}$ or $u_{\infty}$ as the order parameter from the conclusion in Sec. \ref{secPT}. This has some implications for the diversity observed in natural dynamic earthquake slip behavior. For example, the dependence of stress drops on earthquake size can be explained. The stress drop $\Delta \sigma$ is the difference between the applied shear stress acting on the fault plane and the residual frictional stress, as mentioned in Sec. \ref{secMS}. Some researchers insist that large earthquakes sometimes have larger dynamic stress drops than other earthquakes \cite{Dup}. This mechanism can be understood by the framework here. For such large earthquakes, the point attractor (high-speed phase) may be realized because $\Delta \sigma \equiv \sigma_s^0-\sigma_{\mathrm{res}} =\mu v/2 \beta_v=\sigma_s^0 v$ (see Eqs. (\ref{eqEOM}) and (\ref{eqNv})) remains nonzero. This corresponds to the case wherein the acceleration by the fluid pressure increment due to the thermal pressurization effect completely governs the system behavior, and the shear stress acting on the fault plane is completely released owing to thermal pressurization. The fluid pressure approaches $-\sigma_n^0$ in this case. On the other hand, other ordinary earthquakes may be realizations of the line attractor (cessation phase), because we clearly have $\Delta \sigma=0$ in the cessation phase. This behavior corresponds to a situation in which deceleration due to fluid pressure reduction induced by the dilatancy effect completely governs the slip behavior, and spontaneous slip cessation ($v=0$) is realized. Note that the 1D system utilized here is an approximated one, and $\Delta \sigma$ remains nonzero near the fault tips for real three-dimensional systems. However, the near-tip area becomes negligibly small compared with the whole fault area with propagation of the tip, and the 1D approximation is expected to work well for natural faults.

The slip behavior from the onset to the attainment of two phases can be interpreted in terms of two physical processes, the thermal pressurization and dilatancy effects. Note that the dominant physical processes can exchange during the slip. If the solution orbit is on the upside (underside) of $C^{\mathrm{crit}}$, the gradient of the orbit is negative (positive) and $v$ decreases (increases) with increasing time. The deceleration (acceleration) of the slip is physically observed, and the dilatancy (thermal pressurization) effect is dominant. The curve $C^{\mathrm{crit}}$, i.e., the function $g(\phi)$, determines which physical process is dominant. However, it should be emphasized that the high-speed and cessation phases are completely governed by the thermal pressurization and dilatancy effects, respectively, and an intermediate state at $\tau \to \infty$ does not exist, as mentioned above. Although this phase transition was also suggested in SY14, only a single form for the porosity evolution law was assumed, and the order parameter ($\phi_{\infty}$ or $u_{\infty}$) was not clarified there.

The question here is whether we can predict which phase appears based on the physical viewpoint. As noted in Sec. \ref{secPLS}, $\beta$, $v_0$, and $g(\phi)$ govern which phase emerges via the sign of relation (\ref{eqGS}). We can completely predict slip behavior mathematically. Nonetheless, $g(\phi)=1-\beta f(F^{-1}(\phi))=1-\beta f(u) =1-(\beta/\phi_{\mathrm{UL}}) \partial \phi_d (U_{\mathrm{ref}} u)/\partial u$ depends upon the porosity evolution law, which has not been firmly understood, as mentioned in Sec. \ref{secI}. Thus, the law is so uncertain that we cannot predict which phase emerges from the physical viewpoint.

Though predicting which phase emerges is difficult, it is physically meaningful to assume that the cessation phase emerges because earthquakes with enormously large stress drops are rare \cite{Dup}. This observational result implies that $v_0-v_{\mathrm{right}}^1<0$ is satisfied for many earthquakes. The question arising from the assumption of the cessation phase is whether we can predict the final slip amount.  Note here that the value of the final slip amount $u_{\infty}=u_c-\delta u$ is a measure of earthquake magnitude (which cannot be defined exactly here because the model is 1D), and that studying the behavior of $u_{\infty}$ is important for understanding the dynamic earthquake slip process. We adopt the model of SY14 as an example, and perform numerical calculations. We will derive implications independent of $\beta$ and $g(\phi)$.

The governing equations in SY14 are given by
\begin{equation}
\dot{v}=v(1-v)-S_u (1-\phi)v, \label{eqGovSY14-1}
\end{equation}
\begin{equation}
\dot{\phi}=T_a (1-\phi) v, \label{eqGovSY14-2}
\end{equation}
where $S_u$ and $T_a$ are nondimensional positive parameters. Applying the notation in the present study to this equation system, we have $f(F^{-1}(\phi))=T_a (1-\phi)$, $g(\phi)=1-S_u (1-\phi)$, and $\beta=S_u/T_a$. With these conditions and Eqs.  (\ref{eqRelUPhi}), (\ref{eqSolPhic}), and (\ref{eqvc}), we can write $v_c$, $\phi_c$ and $u_c$ for SY14:
\begin{equation}
v_c = 1-\frac{S_u^{1/T_a} T_a - S_u}{T_a-1}, \label{eqSolvcEx}
\end{equation}
\begin{equation}
\phi_c = 1-\frac{1}{S_u}, \label{eqSolPhicEx}
\end{equation}
\begin{equation}
u_c=\frac{\ln S_u}{T_a}, \label{eqSolucEx}
\end{equation}
respectively. Finally, we can apply Eq. (\ref{eqP3}) to the model of SY14 to obtain
\begin{equation}
\delta u = \delta v^{1/2} \sqrt{\frac{2}{S_u^{1/T_a} T_a}} . \label{eqExm1}
\end{equation}
The values of $S_u$ and $T_a$ to be utilized are those which are concluded to be appropriate for ordinary earthquakes in SY14. The Runge-Kutta method with the fourth-order accuracy is adopted. The calculated $u$ was found to approach $u_{\infty}$ within the numerical accuracy in finite time.

\begin{figure*}[tbp]
\centering
\begin{minipage}[t]{8.cm}
\includegraphics[width=8.cm]{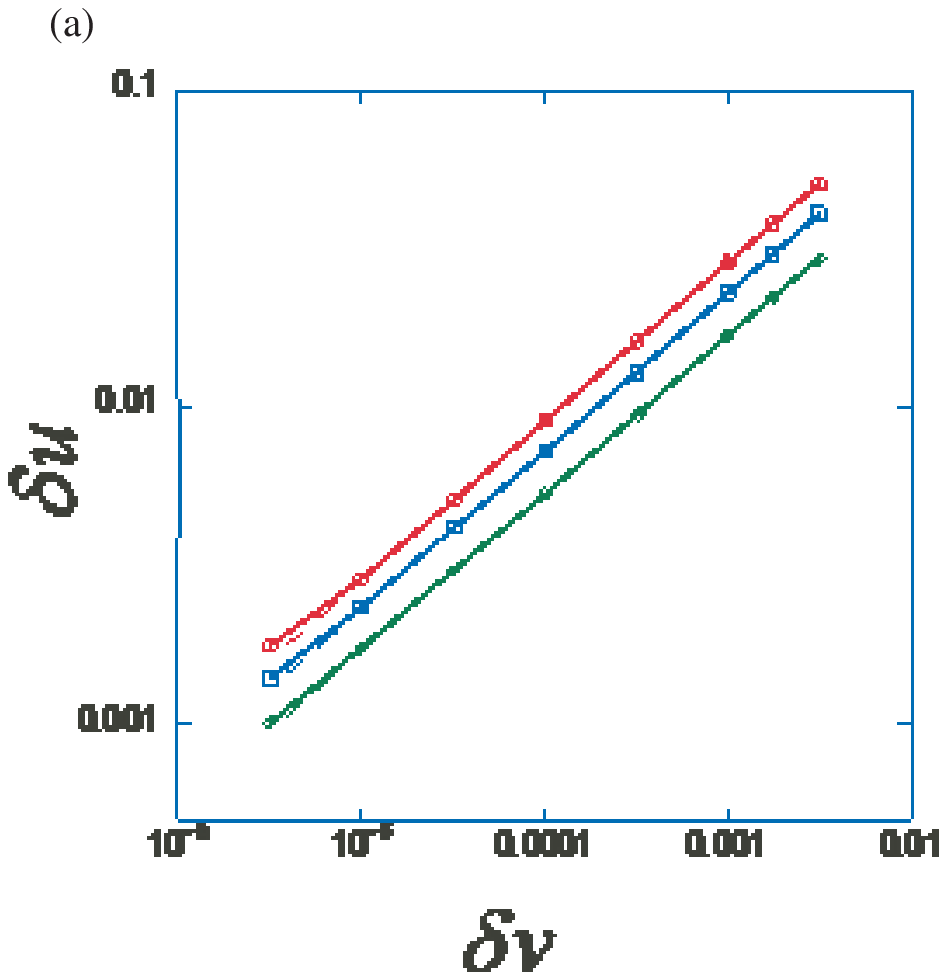}
\end{minipage}
\begin{tabular}{cc}
\begin{minipage}[t]{7.5cm}
\includegraphics[width=7.5cm]{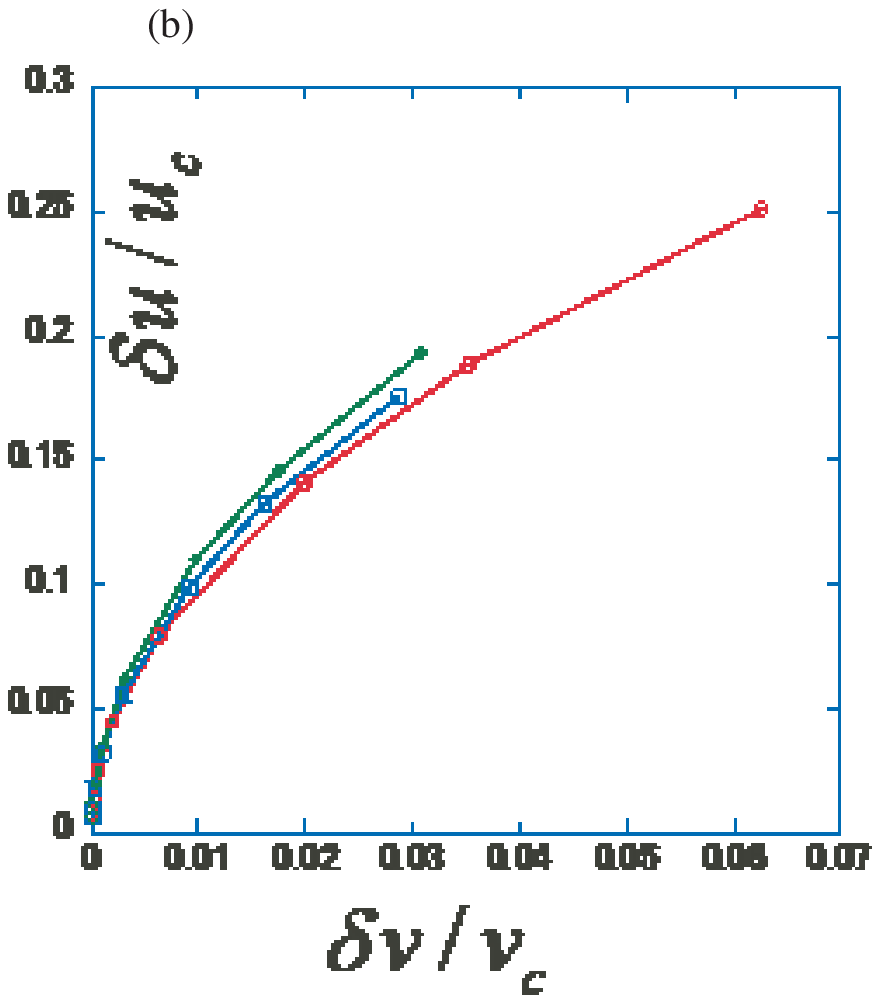}
\end{minipage} &
\begin{minipage}[t]{7.5cm}
\includegraphics[width=7.5cm]{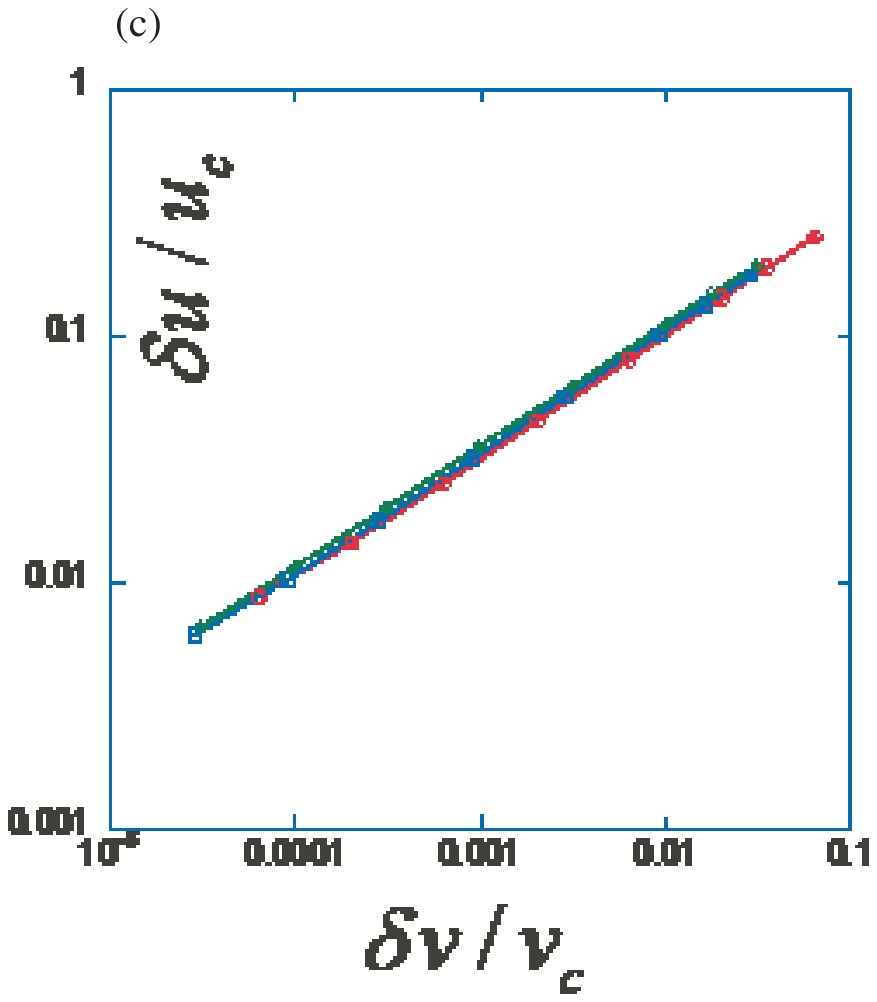}
\end{minipage}
\end{tabular}
\caption{Power law. (a) Relations between $\delta v$ and $\delta u$, and their dependence on the values of $(S_u, T_a)$. The values of $(S_u, T_a)$ are $(1.5, 2)$ (red), $(2, 3)$ (blue) and $(2.5, 6)$ (green). The solid lines are numerically obtained relations, whereas the dotted lines describe Eq. (\ref{eqExm1}). The numerical curves nearly overlap the analytical ones. (b) Relations between $\delta v/v_c$ and $\delta u/u_c$. Values of $v_c$ and $u_c$ are calculated based on Eqs. (\ref{eqSolvcEx}) and (\ref{eqSolucEx}), respectively. (c) Log-log scale version of (b).}
\label{FigPower}
\end{figure*}

Figure \ref{FigPower}(a) shows the numerically obtained relation between $\delta v$ and $\delta u$, which agrees with the result of Eq. (\ref{eqExm1}). In addition, Figs. \ref{FigPower}(b) and (c) show that the relative disturbance in the final slip amount, $\delta u/u_c$, is several to 10 times larger than that in the initial slip velocity, $\delta v/v_c$. This occurs because we have the universal critical exponent 1/2. Since the uncertainty in $\delta v$ is significantly amplified in $\delta u$, predicting $\delta u$ is almost impossible, and thus the final slip amount of the earthquake is hard to predict. This non-predictability is newly suggested here, and was not studied in previous studies including SY14.

Equation (\ref{eqP3}) shows that $\delta u$ is proportional to $\delta v^{1/2}$, and that the factor of proportionality depends on $\beta$ and $g(\phi)$.  Therefore, we should emphasize the non-predictability as a universal conclusion independent of uncertainties of $g(\phi)$, and the non-predictability is not specific to the model of SY14.

The parameter $\delta v=v_c-v_0$ depends on $\beta$, $g(\phi)$ and $v_0$ (see Eq. (\ref{eqGS})), such that it is hard to evaluate it from the physical viewpoint. Moreover, $\delta v$ has another uncertainty. Note that $v_0= 1+\mu_{\mathrm{slid}}(\sigma_n^0+p_{D0})/\sigma_s^0$, where $p_{D0}$ is the initial fluid pressure. Among the parameters $\mu_{\mathrm{slid}}$, $\sigma_s^0$, $\sigma_n^0$, and $p_{D0}$, note that $p_{D0}$ is considered to be the most susceptible to the surrounding environment. The change of the fluid pressure within the fault rocks may be caused by some chemical processes such as the dehydration of hydrous minerals \cite{Dob, Jun}, hence its quantitative evaluation becomes very arduous, because it is a microscopic phenomenon. It can be concluded that the final slip amount may reflect the fluid pressure profile, and that its exact prediction is significantly difficult.

\section{DISCUSSION AND CONCLUSIONS} \label{secDisCon}

The present study provides a unified framework to treat thermal pressurization and dilatancy effects simultaneously. In particular, the nullclines common to the two equations are important in this framework. Such a system generates geometrically different attractors: the point attractor (high-speed phase in a physical sense) and the line attractor (cessation phase). The transition behavior between the two phases is observed near the basin boundary, and the universality in the vicinity of the transition point is represented by the power law between $\delta \phi$ and $\delta v$, regardless of the details of the porosity evolution law: $\delta \phi \propto \delta v^{1/2}$. Dynamic earthquake slip processes can be regarded as phase transition phenomena, with the final inelastic porosity or the final slip amount as the order parameter. The prediction of the emergent phase is completely performed mathematically, whereas predicting it physically is difficult, mainly because the porosity evolution law has uncertainties. Moreover, even if we assume that the cessation phase emerges, the universality suggests the non-predictability of the final slip amounts of natural earthquakes.

If a complete framework for the porosity evolution law could be developed, we may predict the phase emergence by identifying parameters used to construct $v_0=1+\mu_{\mathrm{slid}} (\sigma_n^0+p_{D0})/ \sigma_s^0$,  $\beta=M \mu_{\mathrm{slid}} \phi_{\mathrm{UL}}/\sigma_s^0$ and $U_{\mathrm{ref}}=((1-\phi_t) \rho_s C_s +\phi_t \rho_f C_f)w_h/((b-\phi_t) \alpha_s + \phi_t \alpha_f)M \mu_{\mathrm{slid}}$, because $g(\phi)$ is given by $g(\phi)=1-(\beta/\phi_{\mathrm{UL}}) \partial \phi_d (U_{\mathrm{ref}} u)/\partial u$. A more detailed study of the porosity evolution law is important for understanding the phase emergence. However, it should be noted that as mentioned in Sec. \ref{secApp}, evaluating $v_0$ is difficult. A stochastic approach could eventually make it easier, and the prediction would have a stochastic character. We should also emphasize that even in the case of high probability for the cessation-phase emergence, the final slip amount could not be predicted because of the universality independent of $\beta$ and $g(\phi)$.

Nonetheless, it is meaningful to investigate the effect of $\beta$ because it is the single parameter emerging in the governing equation system. As mentioned in Sec. \ref{secMS}, $\beta$ is a measure of the contribution of the inelastic porosity increase to the slip velocity change. Considering the region $0 \le \phi \le 1$ and $0 \le v \le 1$, and assuming that $f$ is completely understood, it can be found that larger $\beta$ generates larger area where $v > g$ within such a region since $g=1- \beta f$; in other words $C^{\mathrm{crit}}$ shifts downward, but it keeps crossing the point $(1,1)$. Therefore, it can be concluded that larger $\beta$ (e.g., larger $M$ or smaller $\sigma_s^0$) is more likely to generate the cessation phase. This means that if the porosity evolution law is understood, the tendency by which the cessation phase is likely to emerge can be estimated from $\beta$.

We can perform nonlinear mathematical applications of the framework constructed in this paper. In particular, we now discuss the case of $g'(\phi_c^m)=0$ mathematically. Let us assume again that $\forall j, \ g^{(j)} (\phi_c^m)=0$, $g^{(n)}(\phi_c^m) \neq 0$, and $n \ge 2$. Defining $\delta \phi_{(m)} \equiv |\phi_c^m-\phi_{\infty}|$ and $\delta v_{(m)} \equiv |v_c^m-v_0|$, where $v_c^m$ is the $v$ value at which the manifold crossing $(\phi_c^m, 0)$ crosses the $v$-axis, we can show the relation
\begin{equation}
\delta \phi_{(m)} =(\delta v_{(m)})^{1/(n+1)} \Big| \frac{(n+1)! e^{\beta(A(0)-A(\phi_c^m))}}{\beta g^{(n)} (\phi_c^m)} \Big|^{1/(n+1)} \label{eqPn}
\end{equation}
(see details in Appendix \ref{secAg}). In this case, the universal critical power value is $1/(n+1)$, which decreases with increasing $n$. In particular, if $n$ is even and $g^{(n)}(\phi_c^m)<0$, $\phi_c^m-\phi_{\infty}$ and $v_c^m-v_0$ can take both positive and negative values because the point $(\phi_c^m, 0)$ is on a line attractor and not its end point. In this case, Eq. (\ref{eqPn}) predicts that the region on the $\phi$-axis near the point $(\phi_c^m, 0)$ is harder for the solution to approach with larger $n$, because the disturbance in $\delta v_{(m)}$ is enlarged beyond that in $\delta \phi_{(m)}$ (note that $\delta v_{(m)}<1$ and $\delta \phi_{(m)}<1$) and the enlargement is stronger, even though the region is on an attractor. These treatments are important from the viewpoint of nonlinear mathematics.

Finally, we can show that the treatments performed in this study can be extended to systems such as competition relations between two species (the LV model). The competitive LV model is a simple model of the population dynamics of species competing for some common resource. The framework constructed in the present article represents a special case for the system. To show this, first, note that the competitive LV model (in the absence of the diffusion terms) is given by the following equations:
\begin{equation}
\frac{d x_1}{dt}=r_1 x_1 \left( 1-\frac{1}{K_1} \cdot x_1 -\frac{a_{12}}{K_1} \cdot x_2 \right), \label{eqLV1}
\end{equation}
\begin{equation}
\frac{d x_2}{dt}=r_2 x_2 \left( 1-\frac{a_{21}}{K_2} \cdot x_1 -\frac{1}{K_2} \cdot x_2 \right), \label{eqLV2}
\end{equation}
where $x_1$ ($x_2$) is the population size of species 1 (2), $r_1$ ($r_2$) is the inherent per-capita growth rate of 1 (2), $K_1$ ($K_2$) is the carrying capacity of 1 (2), and $a_{12}$ ($a_{21}$) represents the effect that species 2 (1) has on the population of species 1 (2). Let us consider the system under conditions that (A) the growth rate for species 2 is negligibly smaller than that for species 1, and (B) when species 1 consumes species 2, species 1 also dies, e.g., species 2 is poisonous. If we consider the limit $r_2 \to 0$ while maintaining $r_2 a_{21}/K_2$ constant, this system can be depicted and the governing equation system is given by
\begin{equation}
\dot{X}_1=X_1 (1 -X_1) -\frac{a_{12} K_2}{K_1} \cdot X_1 X_2, \label{eqMLV1}
\end{equation}
\begin{equation}
\dot{X}_2=-\frac{r_2 a_{21} K_1}{r_1 K_2} \cdot X_1 X_2, \label{eqMLV2}
\end{equation}
where $X_1 \equiv x_1/K_1$ and $X_2 \equiv x_2/K_2$, and temporal differentiation is performed with respect to $\tau^{LV} \equiv r_1 t$.

The system can be described by the equation system in the same manner as Eqs. (\ref{eqGovG3}) and (\ref{eqGovG4}). If we introduce the variables $v \equiv X_1$ and $\phi \equiv 1-X_2$, Eqs. (\ref{eqMLV1}) and (\ref{eqMLV2}) are exactly the same as Eqs. (\ref{eqGovSY14-1}) and (\ref{eqGovSY14-2}), respectively, by replacing $S_u$ and $T_a$ with $a_{12}K_2/K_1$ and $r_2 a_{21} K_1/r_1 K_2$, respectively. Which species survives is determined by two important values. The first one is $v_0^{LV}$, which is the $v$ value at the point where the solution orbit crossing the point $(\phi_{\mathrm{init}}, v_{\mathrm{init}})$ crosses the $v$-axis on the phase space, where $v_{\mathrm{init}}$ and $\phi_{\mathrm{init}}$ are the initial values for $v$ and $\phi$, respectively (note that $v_{\mathrm{init}}$ and $v_0^{LV}$ are different). The variables $(v, \phi, v_0)=(v_{\mathrm{init}}, \phi_{\mathrm{init}}, v_0^{LV})$ must satisfy Eq. (\ref{eqSolPhi}), and we have the relation
\begin{eqnarray}
v_{\mathrm{init}} = 1 &-&\frac{\frac{a_{12} K_2}{K_1} (1-\phi_{\mathrm{init}})}{1-\frac{r_2 a_{21} K_1}{r_1 K_2}} \nonumber \\
&+& \left( v_0^{LV} -1+\frac{\frac{a_{12} K_2}{K_1}}{1-\frac{r_2 a_{21} K_1}{r_1 K_2}} \right) \nonumber \\
&\times& (1-\phi_{\mathrm{init}})^{r_1 K_2/r_2 a_{21} K_1}.
\end{eqnarray}
Solving this equation for $v_0^{LV}$ gives us
\begin{eqnarray}
v_0^{LV}&=&(1-\phi_{\mathrm{init}})^{-r_1 K_2/r_2 a_{21} K_1} \nonumber \\
&\times& \left( v_{\mathrm{init}}-1+\frac{\frac{a_{12}K_2}{K_1} (1-\phi_{\mathrm{init}})}{1-\frac{r_2 a_{21} K_1}{r_1 K_2}} \right) \nonumber \\
&+&1-\frac{\frac{a_{12}K_2}{K_1}}{1-\frac{r_2 a_{21} K_1}{r_1 K_2}}. \label{eqv0LV}
\end{eqnarray}
The other value is $v_c^{LV}$, which is the $v$ value at the point where the critical manifold crosses the $v$-axis. Replacing $S_u$ and $T_a$ with $a_{12}K_2/K_1$ and $r_2 a_{21} K_1/r_1 K_2$, respectively, in Eq. (\ref{eqSolvcEx}) gives the value, which leads to
\begin{equation}
v_c^{LV} = 1-\frac{\left( \frac{a_{12}K_2}{K_1} \right)^{r_1 K_2/r_2 a_{21} K_1} \frac{r_2 a_{21} K_1}{r_1 K_2} - \frac{a_{12}K_2}{K_1}}{\frac{r_2 a_{21} K_1}{r_1 K_2}-1}. \label{eqvcLV}
\end{equation}
Comparison between $v_0^{LV}$ and $v_c^{LV}$ allows an exact evaluation of which species survives; if $v_0^{LV}>v_c^{LV}$, the species $v$ ($x_1$) survives, whereas if $v_0^{LV}<v_c^{LV}$, the species $\phi$ ($x_2$) survives.


\appendix

\renewcommand{\theequation}{A\arabic{equation}}
\setcounter{equation}{0}

\section{THE ORBIT CROSSING $(0,v_0)$} \label{secAA}

We show here that all the orbits from $(\phi, v)=(0,v_0)$ are absorbed into $(1,1)$, not $(1,0)$, based on Eq. (\ref{eqSolPhi}). We first assume $\displaystyle{ \lim_{\phi \to 1} A(\phi) =\infty }$, i.e., we assume $e^{-\beta A(\phi)} =h(\phi)(1-\phi)^{\delta}$, where $h(\phi)$ is a regular function of $\phi$ with $h(1) \neq 0$, and $\delta$ is a positive constant. With this assumption, we have
\begin{eqnarray}
&\lim_{\phi \to 1}& e^{-\beta A(\phi)} (B(\phi)-B(0)) \nonumber \\
&=& \lim_{\phi \to 1} h(\phi)(1-\phi)^{\delta} \int_0^\phi \frac{d \phi^{\ast}}{h(\phi^{\ast}) (1-\phi^{\ast})^{\delta}} \nonumber \\
&=& \lim_{\phi \to 1} h(\phi)(1-\phi)^{\delta} \int_0^\phi (1-\phi^{\ast})^{-\delta} \sum_{i=0}^{\infty} C_i (1-\phi^{\ast})^i d \phi^{\ast} \nonumber \\
&=& \lim_{\phi \to 1} h(\phi) (1-\phi)^{\delta} \left[ \sum_{i=0}^{\infty} C_i \frac{(1-\phi^{\ast})^{i-\delta+1}}{i-\delta+1} \right]_0^{\phi} \nonumber \\
&=&0.
\end{eqnarray}
We have expanded the function $1/h(\phi^{\ast})$ around $\phi^{\ast}=1$, and $C_i$ is the coefficient of the $i$th order of $1-\phi^{\ast}$ for the expansion. Moreover, the term $e^{-\beta(A(\phi)-A(0))}(v_0-1)$ clearly vanishes with the limit $\phi \to 1$, such that we can conclude that orbit (\ref{eqSolPhi}) always crosses the point $(1,1)$ and does not cross $(1,0)$ when $\displaystyle{ \lim_{\phi \to 1} A(\phi) =\infty }$.

\begin{figure}[tbp]
\centering
\includegraphics[width=8.5cm]{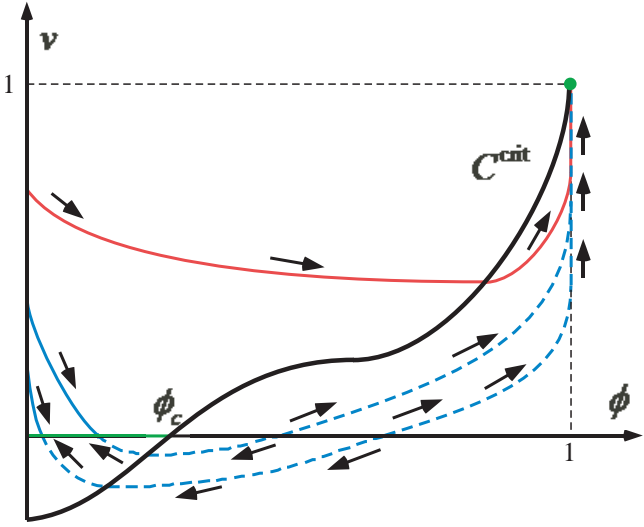}
\caption{The case wherein the solution orbits move vertically on the straight line $\phi=1$. The meanings of the red and blue curves and small arrows are the same as those in Fig. \ref{FigSO}.}
\label{FigSO11}
\end{figure}

We now consider the condition where $\displaystyle{ \lim_{\phi \to 1} A(\phi) }$ does not diverge. In this case, the orbit crosses the point $(\phi, v)=(1,v_1)$, where $v_1$ satisfies the condition $0<v_1<1$. This is because the straight line $\phi=1$ is a nullcline for $\dot{\phi}=0$, and the solution on the line with $v<0$ moves downward vertically with increasing time. The orbit crossing $(1, v_1)$ describes the situation in which the porosity approaches the upper limit ($\phi =1$) within finite time. Then, the orbit ascends the straight line $\phi=1$ vertically, and eventually approaches the point $(1,1)$ (see Fig. \ref{FigSO11}). This describes the case where $C^{\mathrm{crit}}$ is a tangent to the line $\phi=1$, and this case can be included in the investigation performed in this paper. We have shown that there exist no orbits connecting the points $(0,v_0)$ and $(1,0)$ for any forms of $A(\phi)$.

\setcounter{section}{1}
\renewcommand{\thesection}{\Alph{section}}

\renewcommand{\theequation}{B\arabic{equation}}
\setcounter{equation}{0}

\section{THE CONDITION FOR $v_{\mathrm{right}}^1>0$} \label{secAc}

We show here that the condition $v_{\mathrm{right}}^1>0$ is always satisfied when $g(0)<0$. The condition $v_{\mathrm{right}}^1>0$ is equivalent to
\begin{equation}
\beta e^{-\beta A(0)} (B(\phi_{\mathrm{right}}^1)-B(0)) -e^{\beta(A(\phi_{\mathrm{right}}^1)-A(0))}+1 >0,
\end{equation}
which reduces to
\begin{equation}
\int_0^{\phi_{\mathrm{right}}^1} \left(\frac{1}{1-g(\phi_{\mathrm{right}}^1)}-\frac{1}{1-g(\phi)} \right) e^{\beta A(\phi)} d \phi >0.
\end{equation}
This condition is always satisfied if $g(0)<0$ because $g(\phi_{\mathrm{right}}^1)=0$ and $g(\phi) \le 0$ for $0 \le \phi \le \phi_{\mathrm{right}}^1$, which concludes that the integrand is always positive in this case. We can conclude that we always have $v_{\mathrm{right}}^1$ when $g(0)<0$.

\renewcommand{\theequation}{C\arabic{equation}}
\setcounter{equation}{0}

\section{DETAILED TREATMENT OF $\phi_c^m$} \label{secAg}

The cases for $\phi_c^m$ other than $\phi_c^1=\phi_{\mathrm{right}}^1$ are considered here from the mathematical viewpoint. Let $\forall j, \ g^{(j)}(\phi_c^m)=0$ and $g^{(n)}(\phi_c^m) \neq 0$, where $g^{(j)}$ represents the $j$th differentiation of $g$ with respect to $\phi$, $n$ is a positive integer, and $j$ is a nonnegative integer ($0 \le j \le n-1$). We first treat the case $n=1$, $g'(\phi_c^m)>0$ and $\phi_c^m =\phi_{\mathrm{right}}^i$ ($i \ge 2$). In fact, the treatment in this case is exactly the same as that mentioned in Sec. \ref{secPT}, except for the discussion about Eq. (\ref{eqSolPhic}); $g^{-1}(0)$ cannot be defined uniquely and all of $\phi_{\mathrm{right}}^i$ can be written as $g^{-1}(0)$ in this case ($g^{-1}$ is multivalued at $\phi=0$). However, the mathematical treatment is valid even in this case, and the characters $\phi_{\mathrm{right}}^1$, $\delta \phi_+^1$ and $\delta v_+^1$ in Sec. \ref{secPT} can be replaced with $\phi_{\mathrm{right}}^i$, $\delta \phi_+^i \equiv \phi_{\mathrm{right}}^i -\phi_{\infty}$ and $\delta v_+^i \equiv v_{\mathrm{right}}^i -v_0$, respectively. We have the power law $\delta \phi_+^i=(\delta v_+^i)^{1/2} (2 \exp(\beta(A(0)-A(\phi_{\mathrm{right}}^i)))/\beta g'(\phi_{\mathrm{right}}^i))^{1/2}$ near the right end point of the $i$th line attractor.

\begin{figure}[tbp]
\centering
\includegraphics[width=8.5cm]{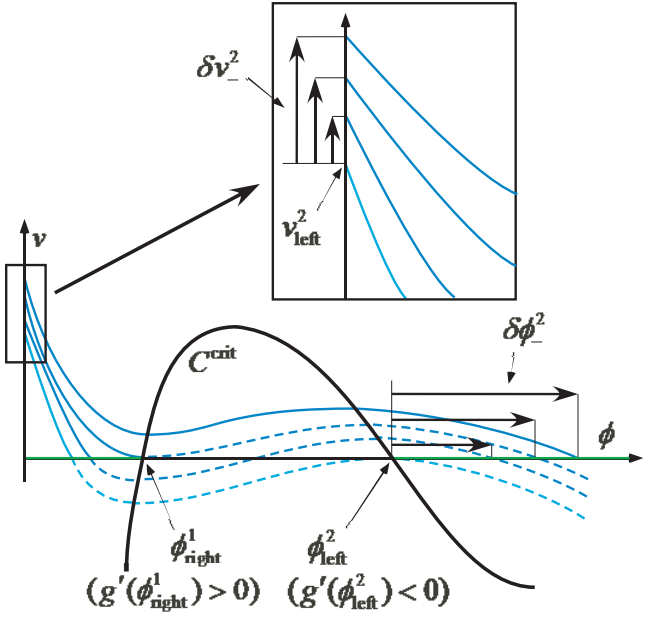}
\caption{Left end point. The solution orbit illustrated by the light blue curve indicates a solution orbit crossing the point $(\phi_{\mathrm{left}}^2, 0)$. The dashed parts of the orbits correspond to mathematical and unphysical solutions.}
\label{FigSOleft}
\end{figure}

We next investigate the case wherein $n=1$ and $g'(\phi_c^m)<0$. In this case $\phi_c^m=\phi_{\mathrm{left}}^i$ ($i \ge 2$). Here, we consider the region near $(\phi_c^2,0)=(\phi_{\mathrm{left}}^2,0)$ (Fig. \ref{FigSOleft}; we do not consider the case wherein the point $(\phi_c^{m_1},0)$ satisfying $g'(\phi_c^{m_1})=0$ appears) as an example. With this assumption, we expect that $v_0$ and $\phi_{\infty}$ for the manifolds absorbed into the line attractor will be larger than $v_{\mathrm{left}}^2$ and $\phi_{\mathrm{left}}^2$, respectively, where $(0, v_{\mathrm{left}}^i)$ is the point where the manifold crossing the point $(\phi_{\mathrm{left}}^i, 0)$ crosses the $v$-axis. In fact, replacing $\delta \phi_+^1$ and $\delta v_+^1$ in the treatment above with $\delta \phi_-^2 \equiv \phi_{\infty}-\phi_{\mathrm{left}}^2$ and $\delta v_-^2 \equiv v_0-v_{\mathrm{left}}^2$, respectively, leads to the power law $\delta \phi_-^2=(\delta v_-^2)^{1/2} (-2 \exp(\beta(A(0)-A(\phi_{\mathrm{left}}^2)))/\beta g'(\phi_{\mathrm{left}}^2))^{1/2}$. This implies that the power law also emerges when $v_0>v_{\mathrm{left}}^2$ and $\phi_{\infty}>\phi_{\mathrm{left}}^2$ for $C^{\mathrm{crit}}$ which is right-downward at $\phi=\phi_{\mathrm{left}}^2$, and the universal power 1/2 appears again. We then consider more than two line attractors as shown in Fig. \ref{FigSOgen}. Even though we have several line attractors in this case, we can apply the investigation above mathematically. At $(\phi_{\mathrm{left}}^i, 0)$, we should use the amounts $\delta \phi_-^i$ and $\delta v_-^i$, which are defined as $\phi_{\infty}-\phi_{\mathrm{left}}^i$ and $v_0-v_{\mathrm{left}}^i$, respectively. The power law can be written as $\delta \phi_-^i=(\delta v_-^i)^{1/2} (-2 \exp(\beta(A(0)-A(\phi_{\mathrm{left}}^i)))/\beta g'(\phi_{\mathrm{left}}^i))^{1/2}$.

\setcounter{section}{3}
\renewcommand{\thesection}{\Alph{section}}

Finally, we treat the case $n \ge 2$. We have $g'(\phi_c^m)=0$, and Eq. (\ref{eqP2}) cannot be adopted in this case, meaning special treatment is required. For this treatment, we first show that
\begin{equation}
\frac{d^l B}{d \phi^l} =\beta^{l-1} \Gamma_{l-1} +\Delta_{l-2} \Gamma_{l-1}, \label{eqLem1}
\end{equation}
where $\Gamma_l =e^{\beta A(\phi)} /(1-g)^l $, $\Delta_{l-2}$ is a polynomial of the form $1-g, g^{(1)}, \cdots, g^{(l-2)}$ and does not include constant terms or those with powers of only $1-g$ (all terms include at least one of the derivatives of $g$), and $l$ is a positive integer satisfying $2 \le l \le n$. We define $\Delta_0 =0$. First, we clearly have the relations
\begin{equation}
\Gamma_l =(1-g) \Gamma_{l+1}
\end{equation}
and
\begin{equation}
\Gamma'_l =\frac{e^{\beta A}}{(1-g)^{l+1}} \beta + e^{\beta A} \cdot (-l) \cdot \frac{-g^{(1)}}{(1-g)^{l+1}} = (\beta+lg^{(1)}) \Gamma_{l+1},
\end{equation}
where the prime denotes the differentiation with respect to $\phi$. To show Eq. (\ref{eqLem1}), we use mathematical induction. We assume that Eq. (\ref{eqLem1}) is satisfied for $l=k$, where $k$ is a positive integer $(\ge 2)$. With this assumption, we have
\begin{eqnarray}
\frac{d^{k+1} B}{d \phi^{k+1}} &=&\beta^{k-1} \Gamma'_{k-1} +\Delta'_{k-2} \Gamma_{k-1} +\Delta_{k-2} \Gamma'_{k-1} \nonumber \\
&=&\beta^{k-1} (\beta+(k-1)g^{(1)}) \Gamma_k +\Delta'_{k-2} (1-g) \Gamma_k \nonumber \\
&+&\Delta_{k-2} (\beta+(k-1)g^{(1)}) \Gamma_k \nonumber \\
&=&\beta^k \Gamma_k +(\beta^{k-1} (k-1)g^{(1)}+\Delta'_{k-2} (1-g) \nonumber \\
&+&\Delta_{k-2} (\beta+(k-1)g^{(1)})) \Gamma_k.
\end{eqnarray}
By defining $\Delta_{k-1} \equiv \beta^{k-1} (k-1) g^{(1)} +\Delta'_{k-2} (1-g) +\Delta_{k-2} (\beta+(k-1)g^{(1)})$, we can see that $\Delta_{k-1}$ is a polynomial in terms of $1-g, g^{(1)}, \cdots, g^{(k-2)}, g^{(k-1)}$ and does not include constant terms or those with powers of only $1-g$. Equation (\ref{eqLem1}) is therefore also satisfied also for $l=k+1$. In addition, we have
\begin{equation}
\frac{d^2 B}{d \phi^2}=\frac{d}{d \phi}e^{\beta A}=\beta \frac{e^{\beta A}}{1-g} =\beta \Gamma_1 +\Delta_0 \Gamma_1,
\end{equation}
such that Eq. (\ref{eqLem1}) is satisfied for $l=2$. We can hence conclude that Eq. (\ref{eqLem1}) is proved by mathematical induction.

Additionally, we show the relation
\begin{equation}
\frac{d^l A}{d \phi^l}=\frac{g^{(l-1)}}{(1-g)^2} +\Theta_{l-2}, \label{eqLem2}
\end{equation}
where $\Theta_{l-2}$ includes positive powers of $g^{(1)}, g^{(2)}, \cdots, g^{(l-2)}$ and negative powers of $1-g$, and does not include constant terms or those with powers of only $1-g$ (all terms include at least one of the derivatives of $g$). We define $\Theta_0=0$. To obtain Eq. (\ref{eqLem2}), we use mathematical induction. We assume that Eq. (\ref{eqLem2}) is satisfied for $l=k$. Under this assumption, we have
\begin{equation}
\frac{d^{k+1} A}{d \phi^{k+1}}=\frac{g^{(k)}}{(1-g)^2}+2 \frac{g^{(k-1)} g^{(1)}}{(1-g)^3} +\Theta'_{k-2}.
\end{equation}
Defining $\Theta_{k-1} \equiv 2 g^{(k-1)} g^{(1)}/(1-g)^3 +\Theta'_{k-2}$, we can see that $\Theta_{k-1}$ includes positive powers of $g^{(1)}, g^{(2)}, \cdots, g^{(k-1)}$ and negative powers of $1-g$, and does not include constant terms or those with powers of only $1-g$. Equation (\ref{eqLem2}) is therefore also satisfied for $l=k+1$. In addition, we have
\begin{equation}
\frac{d^2 A}{d \phi^2}=\frac{g^{(1)}}{(1-g)^2}=\frac{g^{(1)}}{(1-g)^2}+\Theta_0,
\end{equation}
such that Eq. (\ref{eqLem2}) is satisfied for $l=2$. We can thus conclude that Eq. (\ref{eqLem2}) is proved by mathematical induction.

From Eq. (\ref{eqLem1}), we can confirm that
\begin{equation}
\frac{d^l B}{d\phi^l} \Big|_{\phi=\phi_c^m} =\beta^{l-1} \frac{e^{\beta A}}{(1-g)^{l-1}} \Big|_{\phi=\phi_c^m}= \beta^{l-1} e^{\beta A(\phi_c^m)},
\end{equation}
because $g^{(0)}(\phi_c^m)=g^{(1)}(\phi_c^m)=g^{(2)} (\phi_c^m)= \cdots =g^{(l-2)}(\phi_c^m)=0$ and $\Delta_0=0$, which leads to $\Delta_{l-2}|_{\phi=\phi_c^m}=0$. We can rewrite Eq. (\ref{eqEx1}) as
\begin{eqnarray}
\beta \delta \phi_{(m)} &-&\frac{\beta^2}{2} (\delta \phi_{(m)})^2 + \cdots -\frac{\beta^n}{n!} (-\delta \phi_{(m)})^n \nonumber \\
&-&\frac{\beta^{n+1}}{(n+1)!} (-\delta \phi_{(m)})^{n+1} \nonumber \\
&-&1 -e^{\beta(A(0)-A(\phi_c^m))} \delta v_{(m)} +e^{\beta (A(\phi_c^m-\delta \phi_{(m)})-A(\phi_c^m))} \nonumber \\
&+&O((-\delta \phi_{(m)})^{n+2})=0, \label{eqExA1}
\end{eqnarray}
where $\delta \phi_{(m)} \equiv |\phi_c^m-\phi_{\infty}|$ and $\delta v_{(m)} \equiv |v_c^m-v_0|$, and $v_c^m$ is the $v$ value at which the manifold crossing $(\phi_c^m, 0)$ crosses the $v$-axis. We also obtain from Eq. (\ref{eqLem2})
\begin{eqnarray}
A(\phi_c^m-\delta \phi_{(m)})-A(\phi_c^m)=&-&\delta \phi_{(m)} \nonumber \\
&-&\frac{g^{(n)}(\phi_c^m)}{(n+1)!}  (-\delta \phi_{(m)})^{n+1} \nonumber \\
&+&O((-\delta \phi_{(m)})^{n+2}), \label{eqExA2}
\end{eqnarray}
because $g^{(0)}(\phi_c^m)=g^{(1)}(\phi_c^m)=g^{(2)} (\phi_c^m)= \cdots =g^{(l-2)}(\phi_c^m)=0$ and $\Theta_0=0$, which leads to $\Theta_{l-2}|_{\phi=\phi_c^m}=0$. From Eqs. (\ref{eqExA1}) and (\ref{eqExA2}) and by neglecting $O((-\delta \phi_{(m)})^{n+2})$, we obtain the simple power law
\begin{equation}
\delta \phi_{(m)} =(\delta v_{(m)})^{1/(n+1)} \Big| \frac{(n+1)! e^{\beta(A(0)-A(\phi_c^m))}}{\beta g^{(n)} (\phi_c^m)} \Big|^{1/(n+1)},
\end{equation}
because terms lower than $(\delta \phi_{(m)})^n$ vanish. If $n$ is odd and $g^{(n)}(\phi_c^m)<0$, the point $(\phi_c^m, 0)$ is the left end point and $\delta \phi_{(m)} = -(\phi_c^m-\phi_{\infty})$ and $\delta v_{(m)} = -(v_c^m-v_0)$. If $n$ is odd and $g^{(n)}(\phi_c^m)>0$, the point $(\phi_c^m, 0)$ is the right end point and $\delta \phi_{(m)} = \phi_c^m-\phi_{\infty}$ and $\delta v_{(m)} =v_c^m-v_0$. If $n$ is even and $g^{(n)}(\phi_c^m)>0$, a converged line attractor is observed at $(\phi_c^m,0)$ and relation (\ref{eqPn}) does not appear because the region near the point cannot be a part of the line attractor. If $n$ is even and $g^{(n)}(\phi_c^m)<0$, $\phi_c^m-\phi_{\infty}$ and $v_c^m-v_0$ can accept both positive and negative values. We have obtained the relation between $\delta \phi_{(m)}$ and $\delta v_{(m)}$ even in the case of $n \ge 2$.

\acknowledgments
T. S. thanks Prof. T. Mizuguchi, Prof. Y. Sumino, Prof. K. Takeuchi, Prof. K. Ujiie, and Prof. T. Yamashita for fruitful discussions. He was supported by JSPS KAKENHI Grant Numbers JP16K17795 and JP16H06478. This study was supported by the Earthquake Research Institute cooperative research program. The author would like to thank Enago (www.enago.jp) for the English language review.

\end{document}